\documentclass[twocolumn,english,superscriptaddress,citeautoscript,showpacs,preprintnumbers,amsmath,amssymb,prd,floatfix,footinbib]{revtex4}

\usepackage{amsmath, amsthm, amssymb, graphicx,bm,amstext,subfigure}

\begin{document}
\title{Motion of a Mirror under Infinitely Fluctuating Quantum Vacuum Stress}
\author{Qingdi Wang}
\author{William G. Unruh}
\affiliation{Department of Physics and Astronomy, 
The University of British Columbia,
Vancouver, Canada V6T 1Z1}
\date{\today}
\begin{abstract}
The actual value of the quantum vacuum energy density is generally regarded as irrelevant in non-gravitational physics. However, this paper gives a non-gravitational system where this value does have physical significance. The system is a mirror with an internal degree of freedom which interacts with a scalar field. We find that the force exerted on the mirror by the field vacuum undergoes wild fluctuations with a magnitude proportional to the value of the vacuum energy density, which is mathematically infinite. This infinite fluctuating force gives infinite instantaneous acceleration of the mirror. We show that this infinite fluctuating force and infinite instantaneous acceleration make sense because they will not result in infinite fluctuation of the mirror's position. On the contrary, the mirror's fluctuating motion will be confined in a small region due to two special properties of the quantum vacuum: (1) the vacuum friction which resists the mirror's motion and (2) the strong anti-correlation of vacuum fluctuations which constantly changes the direction of the mirror's infinite instantaneous acceleration and thus cancels the effect of infinities to make the fluctuation of the mirror's position finite.
\end{abstract}
\maketitle

\section{Introduction}
In quantum field theory, the vacuum, which is defined as the state of lowest possible energy, is not really empty. Its energy is not zero but infinite since it is associated with the zero-point fluctuations of infinite number of quantum harmonic oscillators. On one hand, it is generally accepted that these zero-point fluctuations really exist in nature \cite{Martin:2012bt} since their physical effects can be experimentally observed in various phenomena such as the spontaneous emission \cite{scully1997quantum}, the Lamb shift \cite{Lamb1947} and the Casimir effect \cite{Casimir:1948dh}. On the other hand, the infinite value of the vacuum energy is generally regarded as irrelevant since experiments measure only energy differences from the ground state. For example, the Casimir effect, which is the small attractive force between two close parallel uncharged conducting plates, happens because the Casimir vacuum energy density decreases as the plates are moved closer, or in other words, it comes from a difference of vacuum energies and in practical calculations the infinities cancel.

Nevertheless, the quantum vacuum never stops astonishing us \cite{Rugh:2000ji}. For example, when it comes to gravity, the actual value of energy matters, not only the difference. According to the principle of General Relativity, the energy momentum tensor is the source of gravitational field. So it is expected that the non-zero vacuum energy will contribute to the cosmological constant, which explains the accelerated expansion of the Universe. Unfortunately, as we stated before, the vacuum energy is mathematically infinite without renormalization and thus would cause a huge cosmological constant for a cut-off at the Planck scale, which disagrees with the tiny measured cosmological constant by a factor of $10^{120}$ \cite{RevModPhys.61.1}. This discrepancy has been called ``the worst theoretical prediction in the history of physics" \cite{hobson2006general}!

It is generally accepted that the actual value of the vacuum energy matters only when taking gravity into account, otherwise one can only measure the energy differences. However, in this paper, we give a non-gravitational physical system where the infinities like that of the vacuum energy do matter. The system is a mirror with an internal harmonic oscillator coupled to a real scalar field in $1+1$ dimension. We find that the fluctuations of the force exerted on the mirror by the field are proportional to the infinite value of the quantum vacuum energy of the scalar field. This infinite fluctuation of force gives infinite instantaneous acceleration of the mirror. However, unlike the vacuum catastrophe in the cosmological constant problem, it is shown that this infinite fluctuating force makes sense because they will not result in infinite fluctuation of the mirror's position. On the contrary, the mirror's fluctuating motion will be confined in a small region due to two special properties of the quantum vacuum: the vacuum friction and the strong anti-correlation of vacuum fluctuations. More precisely, this comes about because (1) there exists vacuum friction (also infinite but with much lower order divergence) to resist the mirror's motion and (2) the force is strongly anti-correlated in time and time average of the force have finite fluctuations. Then although the instantaneous acceleration is infinite, it constantly changes directions, which strongly cancels the effect of infinities and makes the fluctuation of the mirror's position finite. 

This paper is organized as follows. In section \ref{our mirror model}, we introduce our special mirror model and explain how it works in detail. In section \ref{sec force fluctuatioin}, we calculate the force acting on the mirror by the field and its fluctuation. The infinite fluctuations of this force, which are proportional to the value of the vacuum energy, are given. In section \ref{sec: force average}, we calculate the fluctuation of the time average of the force, and find a finite result, which is an indication that our mirror's fluctuating motion under the infinitely fluctuating force might be finite. In section \ref{sec:Motion effects of the mirror}, we examine the frictional force acting on the mirror due to radiation reaction. In section \ref{sec: mirror equation of motion}, we derive the mirror's equation of motion. In section \ref{sec: confined motion}, we calculate the fluctuating motion of the mirror and show it is confined to a small region. In section \ref{sec: discussions}, we compare our mirror's fluctuating motion with Brownian motion and indicate the intrinsic differences between them. In section \ref{sec: conclusions}, we discuss our results and compare them with other related works.

Units are chosen throughout such that $c=\hbar=1$.

\section{Our mirror model}\label{our mirror model}
A mirror is an object that reflects light. In the classical electrodynamics, light waves incident on a material induce small oscillations of the individual particles, for example, electrons in glass, causing each particle to radiate a small secondary wave. All these waves add up together to give reflected and refracted waves. We shall study the case of a mirror that interacting with a massless scalar field. One often uses a perfectly reflecting boundary as a mirror model, i.e. the mirror reflects all wave modes with arbitrarily high frequencies, by imposing the boundary condition that the scalar field vanishes on the surface of the mirror (Fulling and Davies \cite{Fulling:1976yv}, Eq.(2.3); Berrell and Davies \cite{birrell1984quantum}, Eq.(4.43)):
\begin{equation}\label{perfect mirror model}
\phi[t,X(t)]=0,
\end{equation}
where $X(t)$ is the trajectory of the mirror. However, a realistic mirror becomes transparent gradually for high frequency wave modes. Some authors \cite{Gour:1998my, Jaekel:1992ef} add an artificial frequency cut-off by assuming that modes of the quantum field $\phi$ with frequencies higher than a specific value is unaffected by the mirror. In this paper, we will not adopt this model. Instead, we will adopt a mirror model in which the transparency for high frequency wave modes appears in a natural way. 

In our model, the oscillating particle inside the mirror is an harmonic oscillator with natural frequency $\Omega$. We consider a $1+1$ dimensional static mirror with an internal dynamic degree of freedom $q$ coupled to a scalar field $\phi$. The mirror is located at position $x=0$ in the space of the scalar field. The total action is given by
\begin{equation}
\label{static mirror action}
\begin{split}
S&=\frac{1}{2}\iint\left(\left(\frac{\partial\phi}{\partial t}\right)^2-\left(\frac{\partial\phi}{\partial x}\right)^2\right)dtdx\\
&+\frac{1}{2}\int\left(\left(\frac{dq}{dt}\right)^2-\Omega^2 q^2\right)dt\\
&+\epsilon\int\frac{d\phi(t,0)}{dt}q(t)dt,
\end{split}
\end{equation}
where $\epsilon$ is the coupling constant. Here it is necessary to point out that the harmonic oscillator $q$ is not oscillating ``in space", it is an ``internal'' degree of freedom, i.e. a 0-dimensional quantum field inside the mirror.

Varying the action \eqref{static mirror action} with respect to $\phi$ and $q$ leads to the Heisenberg equations of motion for the field $\phi$ and the internal degree of freedom $q$:
\begin{equation}\label{field equation}
\ddot{\phi}-\phi''=-\epsilon\dot{q}\delta(x),
\end{equation}
\begin{equation}\label{oscillator equation}
\ddot{q}+\Omega^2 q=\epsilon\dot{\phi}(t,0),
\end{equation}
where the dot $\dot{}$ denotes the time derivative and the prime $'$ the spatial derivative. The solution of \eqref{field equation} is of the following form
\begin{equation}\label{solution}
\phi(t,x)=\phi_0(t,x)-\frac{\epsilon}{2}q(t-|x|),
\end{equation}
where $\phi_0(t,x)$ is the solution of the homogeneous equation
\begin{equation}\label{homogeneous field equation}
\ddot{\phi_0}-\phi_0''=0.
\end{equation}
One can easily check \eqref{solution} is the solution by noticing that
\begin{equation}
\begin{split}
q''(t-|x|)=&-\frac{d}{dx}\left(\dot{q}(t-|x|)sgn(x)\right)\\
=&\ddot{q}(t-|x|)-2\dot{q}(t)\delta(x),
\end{split}
\end{equation}
where the sign function $sgn(x)$ is defined as
\[
  sgn(x) = \left\{
  \begin{array}{l l l}
    -1 & \quad \text{if $x<0$}\\
    1 & \quad \text{if $x>0$}\\
  \end{array} \right.
\]
Substituting (\ref{solution}) into the equation of motion for the internal oscillator (\ref{oscillator equation}) gives
\begin{equation}\label{damped oscillator equation}
\ddot{q}+\frac{\epsilon^2}{2}\dot{q}+\Omega^2 q=\epsilon\dot{\phi_0}(t,0).
\end{equation}
This is exactly an equation of motion for a driven damped harmonic oscillator with natural frequency $\Omega$, damping coefficient $\frac{\epsilon^2}{2}$ and driving force $\epsilon\dot{\phi_0}(t,0)$.

\begin{figure}
\centering
\includegraphics[scale=0.4]{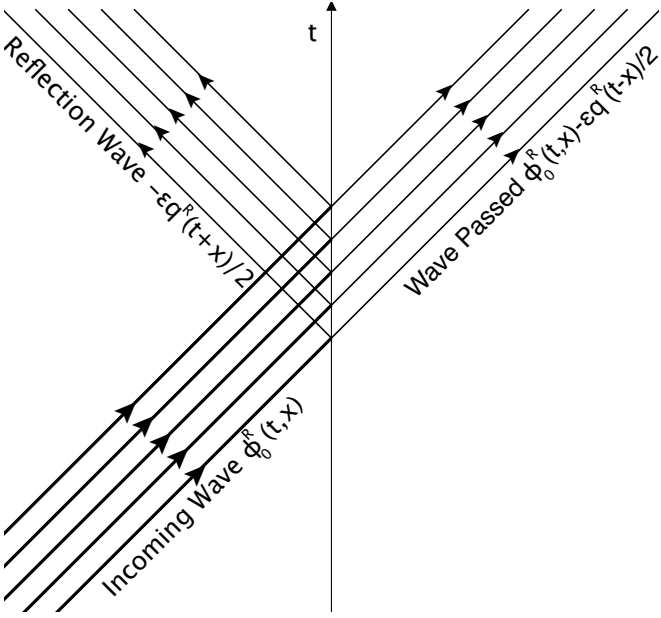}
\caption{\label{mirror working mechanism} A figure shows how our mirror works for right moving wave $\phi_0^R$. When incidents on the mirror, it induces oscillations of the internal harmonic oscillator. Then the oscillator radiates a secondary wave $q^R(t-|x|)$ to both directions equally. For the left moving  wave $\phi_0^L$, the mirror works exactly the same way due to symmetry.}
\end{figure}

In order to give a clear picture about how the mirror works, we divide the incoming field $\phi_0$ into right moving part and left moving part:
\begin{equation}\label{filed decomposition}
\phi_0=\phi_0^R+\phi_0^L,
\end{equation}
where $\phi_0^R$ is the form of $f(t-x)$ and $\phi_0^L$ is the form of $g(t+x)$ according to d'Alembert's solution. This solution has the properties:
\begin{eqnarray}
\dot{\phi}_0^R&=&-{\phi'}_0^R, \label{phir derivative}\\
\dot{\phi}_0^L&=&{\phi'}_0^L, \label{phil derivative}
\end{eqnarray}
which are useful in our later calculations. Since \eqref{field equation} and \eqref{oscillator equation} are liner equations, the internal degree of freedom $q$ can also be divided into two parts correspondingly:
\begin{equation}
q=q^R+q^L,
\end{equation}
and the pairs $(\phi_0^R,q^R)$ and $(\phi_0^L,q^L)$ both obey the same equations \eqref{homogeneous field equation} and \eqref{damped oscillator equation}. The solution \eqref{solution} gives us a picture about how the mirror reflects waves. As shown in FIG. \ref{mirror working mechanism}, the right moving wave $\phi_0^R(t,x)$ is incident on the mirror from left. The mirror reflects a wave $-\frac{\epsilon}{2}q^R(t+x)$ to the left and lets a wave $\phi_0^R-\frac{\epsilon}{2}q^R(t-x)$ pass through to the right. The mirror reflects the left moving wave $\phi_0^L$ in exactly the same way by just doing a ``mirror reflection'' in FIG. \ref{mirror working mechanism}.

\begin{figure}
\centering
\includegraphics[scale=0.3]{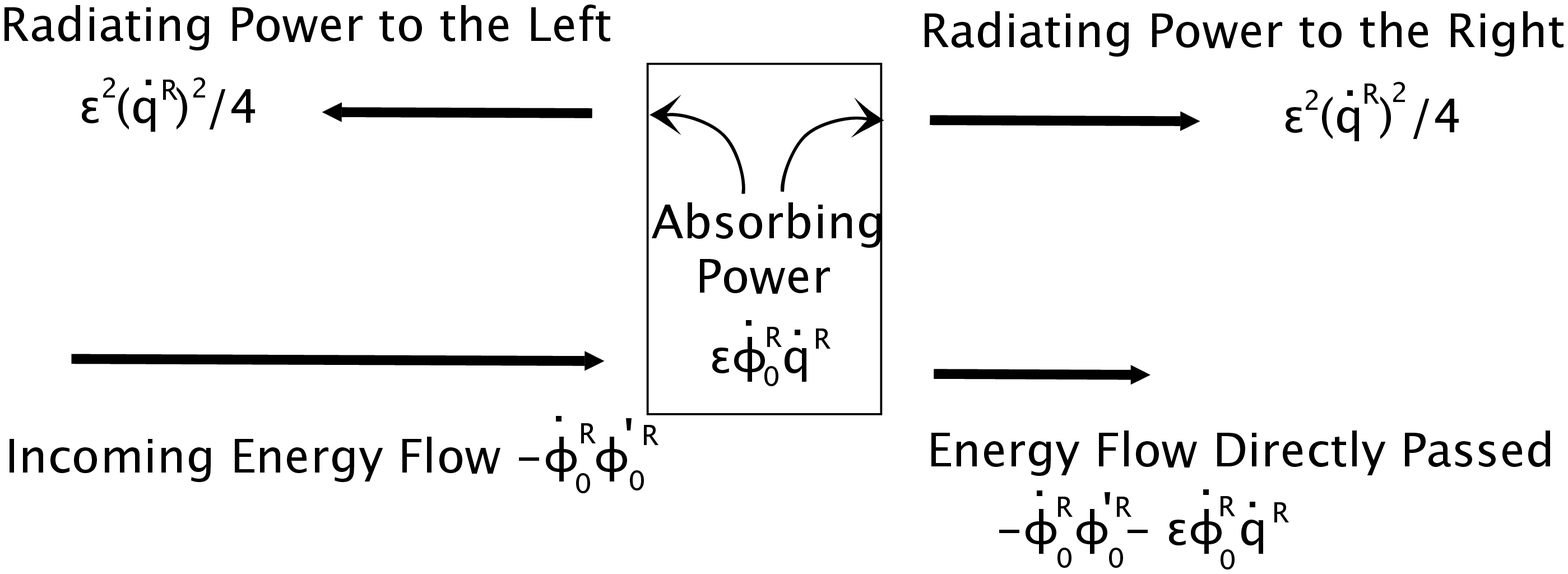}
\caption{\label{reflection process} A figure shows that the incoming filed $\phi_0$ is impinging on the mirror. Part of the field energy is absorbed by the internal harmonic oscillator with absorbing power $\epsilon\dot{\phi}_0^R\dot{q}^R$. At the same time, it is radiating energy out with total power $\frac{\epsilon^2}{2}(\dot{q}^R)^2$.}
\end{figure}

Next, to understand the working mechanism of the mirror in detail, we analyse the energy flows during the reflection process by using the following formula for energy flux:
\begin{equation}\label{energy flux}
T^{01}(t,x)=-\{\dot{\phi}(t,x)\phi'(t,x)\},
\end{equation}
where $T^{01}$ is the time-space component of the type $(2,0)$ stress-energy tensor of the field $\phi$ and the curly bracket $\{\}$ represents the symmetrization operation which is defined as
\begin{equation}\label{symmetric operation}
\{AB\}=\frac{1}{2}(AB+BA),
\end{equation}
for any two operators $A$ and $B$. This is irrelevant for classical quantities but will be important later for quantum operators. For simplicity, we first consider the case that only the right moving wave $\phi_0^R$ exists. Then the energy flux near the left side of the mirror is:
\begin{equation}\begin{split}\label{left side energy flux}
\lim_{x\to 0^-}&\{-\left(\phi_0^R(t,x)-\frac{\epsilon}{2}q^R(t+x)\right)^{\cdot}\\
&\cdot\left(\phi_0^R(t,x)-\frac{\epsilon}{2}q^R(t+x)\right)'\}\\
=&\{\left(-\dot{\phi}_0^R{\phi'}_0^R(t,0)\right)-\left(\frac{\epsilon^2}{4}(\dot{q}^R)^2(t)\right)\},
\end{split}\end{equation}
where we have used \eqref{phir derivative} to eliminate the interference terms. The energy flux near the right side of the mirror is:
\begin{equation}\begin{split}\label{right side energy flux}
\lim_{x\to 0^+}&\{-\left(\phi_0^R(t,x)-\frac{\epsilon}{2}q^R(t-x)\right)^{\cdot}\\
&\cdot\left(\phi_0^R(t,x)-\frac{\epsilon}{2}q^R(t-x)\right)'\}\\
=&\{\left(-\dot{\phi}_0^R{\phi'}_0^R(t,0)-\epsilon\dot{\phi}_0^R(t,0)\dot{q}^R(t)\right)\\
&+\left(\frac{\epsilon^2}{4}(\dot{q}^R)^2(t)\right)\},
\end{split}\end{equation}
where we have again used \eqref{phir derivative}. The first term $-\dot{\phi}_0^R{\phi'}_0^R$ inside the parentheses of \eqref{left side energy flux} represents energy which impinges on the mirror from the left per unit time. The first term $-\dot{\phi}_0^R{\phi'}_0^R-\epsilon\dot{\phi}_0^R\dot{q}^R$ inside the parentheses of \eqref{right side energy flux} represents energy which directly passed through the mirror per unit time. The second term $\frac{\epsilon^2}{4}(\dot{q}^R)^2$ inside the parentheses of \eqref{left side energy flux} represents energy radiated to the left per unit time by the internal harmonic oscillator, which creates the reflective power of the mirror. The same term $\frac{\epsilon^2}{4}(\dot{q}^R)^2$ inside the parentheses of \eqref{right side energy flux} represents energy radiated to the right per unit time by the internal harmonic oscillator. The radiated energy to the left and to the right per unit time add together to give the total radiating power $\frac{\epsilon^2}{2}(\dot{q}^R)^2$. This radiating power is pumped from the incoming wave $\phi_0^R$ with pumping power $\epsilon\dot{\phi}_0^R\dot{q}^R$, which is just the difference between the incoming energy flux $-\dot{\phi}_0^R{\phi'}_0^R$ and the energy flux directly passed through the mirror $-\dot{\phi}_0^R{\phi'}_0^R-\epsilon\dot{\phi}_0^R\dot{q}^R$. This reflection process is illustrated in FIG. \ref{reflection process}.

The reason why the mirror works this way is clear. In fact, note that the internal harmonic oscillator behaves according to the driven damped harmonic oscillation equation \eqref{damped oscillator equation}. From this equation, we notice that the pumping power $\epsilon\dot{\phi}_0^R\dot{q}^R$, which is the ``driving force'' $\epsilon\dot{\phi}_0^R$ times the ``velocity'' $\dot{q}^R$, is exactly the absorbing power from the external driving force. This absorbing power is dissipating by the damping force due to the radiation. The dissipated power is the ``damping force'' $\frac{\epsilon^2}{2}\dot{q}^R$ times the ``velocity'' $\dot{q}^R$, which is exactly equal to the total radiating power $\frac{\epsilon^2}{2}(\dot{q}^R)^2$. So the energy radiated acts as damping on the internal harmonic oscillator. 

In summary, the working mechanism of the mirror is that when the wave incidents on the mirror, part of its energy is used to drive the oscillations of the internal harmonic oscillator; at the same time, the internal harmonic oscillator radiates the absorbed energy out equally to both directions. That energy radiated back forms the reflected waves.

\begin{figure}
\centering
\includegraphics[scale=0.25]{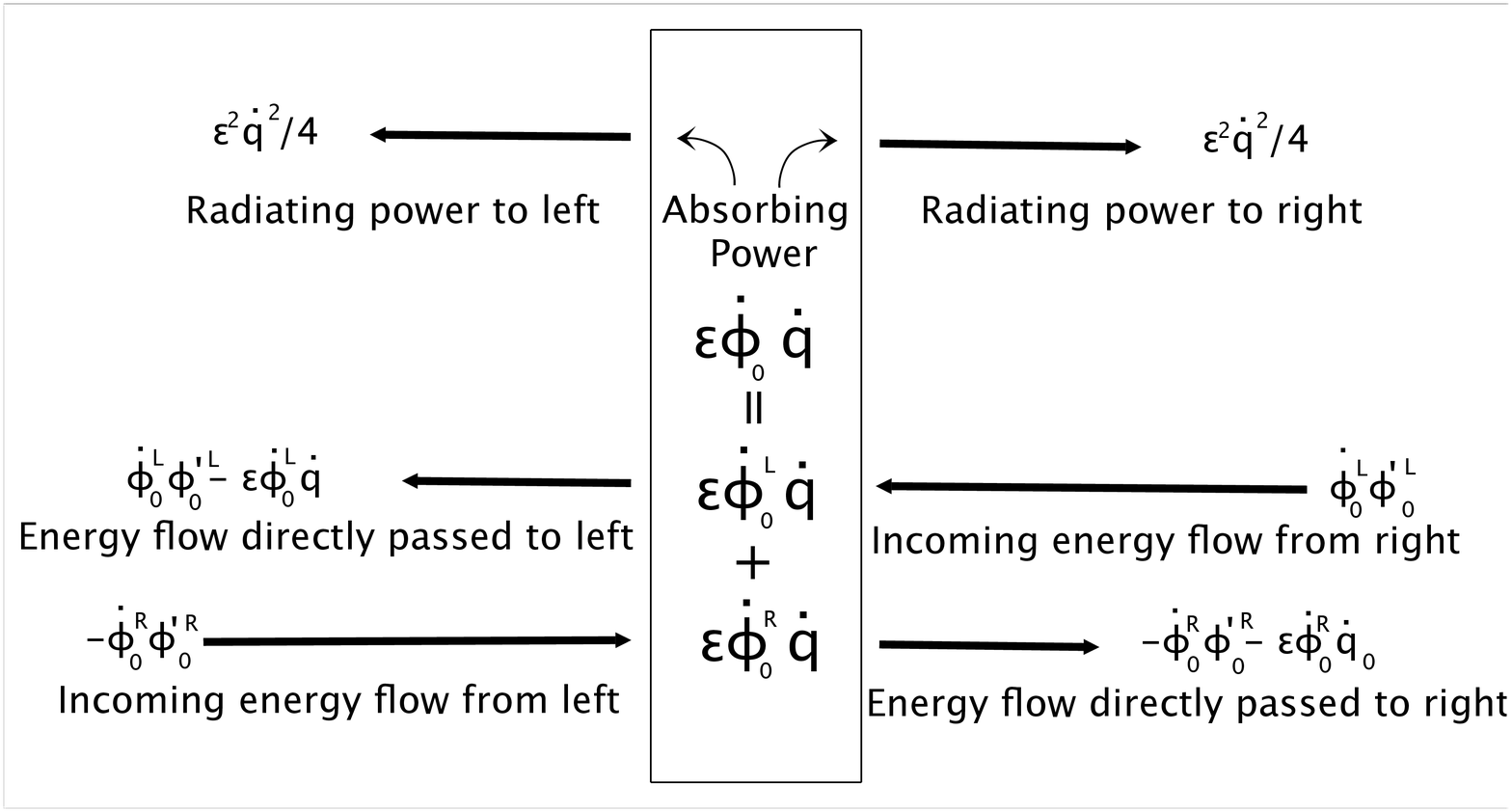}
\caption{\label{mirror energy transfer process} A figure shows that the incoming field $\phi_0$ is impinging on the mirror from both directions. Part of the field energy is absorbed by the internal harmonic oscillator with total absorbing power $\epsilon\dot{\phi}_0\dot{q}$. At the same time, it is radiating energy out with total power $\frac{\epsilon^2}{2}\dot{q}^2$.}
\end{figure}

The mirror works the same way when considering the incoming field $\phi_0$ contains both the right moving part $\phi_0^R$ and the left moving part $\phi_0^L$. Similar calculations show that the energy flux near the left side of the mirror is:
\begin{equation}\begin{split}
&\{\left(-\dot{\phi}_0^R{\phi'}_0^R(t,0)\right)\\
-&\left(\dot{\phi}_0^L{\phi'}_0^L(t,0)-\epsilon\dot{\phi}_0^L(t,0)\dot{q}(t)\right)-\left(\frac{\epsilon^2}{4}\dot{q}^2(t)\right)\}.
\end{split}
\end{equation}
The energy flux near the right side of the mirror is:
\begin{equation}\begin{split}
&\{-\left(\dot{\phi}_0^L{\phi'}_0^L(t,0)\right)\\
+&\left(-\dot{\phi}_0^R{\phi'}_0^R(t,0)-\epsilon\dot{\phi}_0^R(t,0)\dot{q}(t)\right)+\left(\frac{\epsilon^2}{4}\dot{q}^2(t)\right)\}.
\end{split}\end{equation}
The interpretations of the above expressions are similar. We illustrate them in FIG. \ref{mirror energy transfer process}.

Next let us quantize the mirror system. We first go back to the equation of motion \eqref{damped oscillator equation} to analyse the motion of the internal harmonic oscillator $q$ in detail. In our model, the mirror started to interact with the scalar field since $t=-\infty$. So $q$'s initial oscillation has been completely dissipated due to the friction term $\frac{\epsilon^2}{2}\dot{q}$ and the solution of \eqref{damped oscillator equation} is then fully determined by the driving force $\epsilon\dot{\phi_0}(t,0)$:
\begin{equation}\label{static oscillator equation solution}
q(t)=\frac{1}{\omega}\int_{-\infty}^{t}e^{-a(t-t')}\sin(\omega(t-t'))\epsilon\dot{\phi_0}(t',0)dt',
\end{equation}
where $a=\frac{\epsilon^2}{4}$ and $\omega=\sqrt{\Omega^2-\frac{\epsilon^4}{16}}$ is the damped angular frequency. We can quantize the mirror system by expanding $\phi_0$ in terms of the sum of standard annihilation and creation operators $a_k$ and $a_k^{\dag}$:
\begin{equation}\label{field expansion}
\begin{split}
\phi_0(t,x)=\int_{-\infty}^{+\infty}\frac{dk}{\sqrt{4\pi|k|}}\Big(&a_ke^{-i\left(|k|t-kx\right)}\\
+&a_k^{\dag}e^{i\left(|k|t-kx\right)}\Big),
\end{split}
\end{equation}
where the integration over $k$ from $-\infty$ to $0$ represents the left moving modes $\phi_0^L$ and from $0$ to $+\infty$ the right moving modes $\phi_0^R$. Inserting the above expansion \eqref{field expansion} into \eqref{static oscillator equation solution} gives:
\begin{equation}\label{oscillator equation solution}\begin{split}
q(t)=-i\epsilon\int_{-\infty}^{+\infty}&\sqrt{\frac{|k|}{4\pi}}\Bigg(\frac{a_ke^{-i|k|t}}{-k^2-\frac{i}{2}\epsilon^2|k|+\Omega^2}\\
-&\frac{a^{\dag}_ke^{i|k|t}}{-k^2+\frac{i}{2}\epsilon^2|k|+\Omega^2}\Bigg)dk.
\end{split}\end{equation}

\begin{figure}
\centering
\includegraphics[scale=0.4]{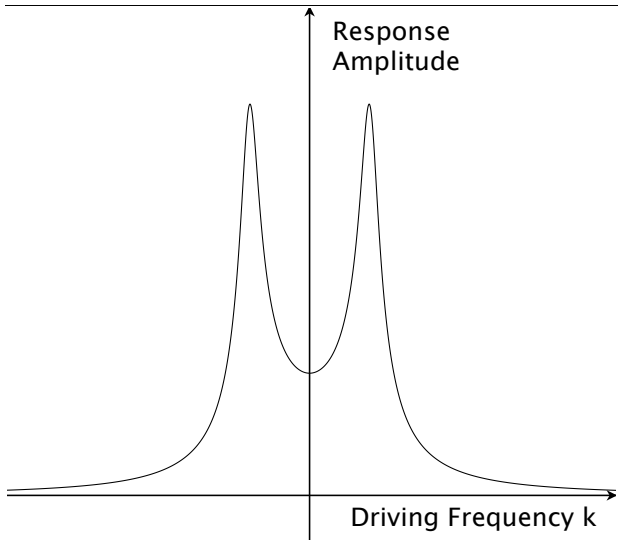}
\caption{\label{spectrum}Steady state variation of amplitude with driving frequency. This graph shows that the internal harmonic oscillator almost not responds to high frequency driven modes.}
\end{figure}

If we evaluate the average radiating power $\left\langle\frac{\epsilon^2}{2}\dot{q}^2\right\rangle$ between frequencies $k$ and $k+\Delta k$ when the system is in vacuum state, which is defined as
\begin{equation}\label{vacuum state definition}
a_k\vert0\rangle=0,
\end{equation}
for any $k\in(-\infty,+\infty)$, we can see that
\begin{equation}
\left\langle p(k)\right\rangle\Delta k=\frac{\epsilon^4}{4\pi}\frac{k^3}{(k^2-\Omega^2)^2+\frac{\epsilon^4}{4}k^2}\Delta k\to 0,
\end{equation}
as $k\to +\infty$. Thus our mirror becomes transparent for high frequency modes. To see more clearly how this transparency property appears, we substitute the annihilation and creation operators $a_k$ and $a_k^{\dag}$ in \eqref{field expansion} by the position and momentum operators $x_k$ and $p_k$:
\begin{equation}
a_k=\sqrt{\frac{|k|}{2}}(x_k+i\frac{p_k}{|k|}),\quad a_k^{\dag}=\sqrt{\frac{|k|}{2}}(x_k-i\frac{p_k}{|k|}).
\end{equation}
Then the driving force can be expressed as
\begin{equation}\label{driven force expansion}
\begin{split}
\epsilon\dot{\phi}_0(t,0)=&\epsilon\int_{-\infty}^{+\infty}\frac{dk}{\sqrt{2\pi}}|k|\\
&\cdot\left(-x_k\sin(|k|t)+\frac{p_k}{|k|}\cos(|k|t)\right),
\end{split}
\end{equation}
which is a sum of infinite number of harmonic oscillation modes with different angular frequencies $|k|$. Plugging \eqref{driven force expansion} into \eqref{static oscillator equation solution} we see that each such mode with a specific frequency $|k|$ drives the motion of $q$ independently since there are no correlations between them. Driven by these independent incoming modes, the damped harmonic oscillator $q$ would be excited and eventually settled down to a steady oscillation state which is also a sum of infinite number of harmonic oscillations with different frequencies and amplitudes:
\begin{equation}\label{internal degree expansion}
\begin{split}
&q(t)=\epsilon\int_{-\infty}^{+\infty}\frac{dk}{\sqrt{2\pi}}\frac{1}{\left[(\Omega^2-k^2)^2+\frac{\epsilon^4}{4}k^2\right]^{1/2}} |k|\\
&\cdot\left(-x_k \sin (|k|t-\alpha_k)+\frac{p_k}{|k|}\cos(|k|t-\alpha_k)\right),
\end{split}
\end{equation}
where $\alpha_k=\arctan(\frac{\epsilon^2|k|}{2(\Omega^2-k^2)})$ is the phase lag. Comparing the integrands of the driving force \eqref{driven force expansion} and the internal driven damped harmonic oscillator \eqref{internal degree expansion} we observe that except for the phase lag $\alpha_k$, the only difference is the factor $\frac{1}{\left[(\Omega^2-k^2)^2+\frac{\epsilon^4}{4}k^2\right]^{1/2}}$ in the latter expression. This factor shows how the mirror becomes transparent for high frequency wave modes. It is just the amplitude response of a damped harmonic oscillator with natural frequency $\Omega$ and damping coefficient $\frac{\epsilon^2}{2}$ driven by a unit oscillating force with frequency $k$ when it reaches the final steady state. As shown in FIG. \ref{spectrum}, the internal oscillator has almost no response for high frequency driving modes. It is this insensitivity that causes the mirror's transparency for high frequency wave modes.

\section{The force on the mirror and its infinite fluctuation} \label{sec force fluctuatioin}
In last section, we introduced our mirror model which is transparent for high frequency wave modes. In this section we will study the force acting on the mirror and its fluctuation when the system is in the vacuum state which is defined in \eqref{vacuum state definition}.

Due to quantum fluctuations, there is net fluctuating force acting on the mirror by the field. As we see from the last section, when the left moving field $\phi_0^L$ and right moving field $\phi_0^R$ incident on the mirror, part of the field energy are absorbed by the internal harmonic oscillator and then the oscillator radiates them out. In this process, the mirror is receiving and sending momentum. On average, the momentum received and sent are symmetric for both sides and thus there is no net force acting on the mirror. However, due to quantum fluctuations, the symmetry between left and right sides can be broken. In other words, sometimes the mirror can receive more momentum from one side than from the other, which gives a net force on the mirror.

The standard definition of the force is:
\begin{equation}\label{force definition}
F(t)=\lim_{x \to 0^+}\left(T^{11}(x_-)-T^{11}(x_+)\right)
\end{equation}
where $x_+=(t,x)$ and $x_-=(t,-x)$ ($x\geqslant0$) are two spacetime points which are symmetrically located on the two sides of the mirror and $T^{11}$ is the space-space component of stress-energy tensor of type $(2,0)$ of the field $\phi$:
\begin{equation}\label{stress energy tensor}
T^{11}(t,x)=\frac{1}{2}\left(\dot{\phi}^2(t,x)+\phi'^2(t,x)\right).
\end{equation}
From \eqref{solution}, we can get the time and space derivatives of the field $\phi$
\begin{equation}\label{phi time derivative}
\dot{\phi}(t,x)=\dot{\phi_0}(t,x)-\frac{\epsilon}{2}\dot{q}(t-|x|),
\end{equation}
\begin{equation}\label{phi spatial derivative}
\phi'(t,x)=\phi_0'(t,x)+\frac{\epsilon}{2}\dot{q}(t-|x|)sgn(x).
\end{equation}
Inserting (\ref{stress energy tensor}), (\ref{phi time derivative}), (\ref{phi spatial derivative}) into (\ref{force definition}) and noticing that when $x$ approaches to $0$, terms of $\dot{\phi}^2(x_-)-\dot{\phi}^2(x_+)$ go to zero due to continuity of $\dot{\phi}$, we have
\begin{equation}\label{force}
F(t)=\{-\epsilon\phi_0'(t,0)\dot{q}(t)\},
\end{equation}
where the curly bracket $\{\}$ is the symmetrization operation defined in \eqref{symmetric operation}. This result is easy to understand. In fact, decomposing $\phi_0$ in the above expression by the sum of left moving modes $\phi_0^L$ and the right moving modes $\phi_0^R$ \eqref{filed decomposition}, and using the properties \eqref{phir derivative} and \eqref{phil derivative}, the above expression becomes
\begin{equation}\label{momentum force}
F(t)=\{\epsilon\dot{\phi}_0^R(t,0)\dot{q}(t)-\epsilon\dot{\phi}_0^L(t,0)\dot{q}(t)\}.
\end{equation}
From FIG. \ref{mirror energy transfer process} we know that $\epsilon\dot{\phi}_0^R\dot{q}$ and $\epsilon\dot{\phi}_0^L\dot{q}$ are energy absorbed per unit time by the mirror from left and from right respectively. Since the field $\phi_0$ is massless, the energy and momentum are the same up to a sign. Thus the above formula is just a manifestation that the force is a sum of momenta absorbed but is also a difference between the energy absorbed from two directions.

From the field expansion (\ref{field expansion}) and the solution of the damped oscillator (\ref{oscillator equation solution}), it is easy to get
\begin{equation}\label{phi'q'}\begin{split}
\left\langle\phi_0'(t,0)\dot{q}(t)\right\rangle&=-i\frac{\epsilon}{4\pi}\int_{-\infty}^{+\infty}\frac{k|k|}{-k^2+\frac{i}{2}\epsilon^2|k|+\Omega^2}dk\\
&=0.
\end{split}
\end{equation}
Thus the expectation value of the force
\begin{equation}
\left\langle F(t)\right\rangle\equiv 0.
\end{equation}
This result is what we expected due to the symmetry of the scalar field: on average, the mirror absorbs equal amount of momentum from both sides. Next we calculate the fluctuation of this force in the vacuum state which is defined as
\begin{equation}
\sigma_F(t)=\left\langle F^2(t)\right\rangle-\left\langle F(t)\right\rangle^2.
\end{equation}
Inserting (\ref{force}) into the above equation gives
\begin{equation}\label{force fluctuation before expansion}
\begin{split}
\sigma_F(t)=&\frac{\epsilon^2}{4}\Big(\left\langle\phi_0'(t,0)\dot{q}(t)\phi_0'(t,0)\dot{q}(t)\right\rangle\\
+&\left\langle\phi_0'(t,0)\dot{q}(t)\dot{q}(t)\phi_0'(t,0)\right\rangle\\
+&\left\langle\dot{q}(t)\phi_0'(t,0)\phi_0'(t,0)\dot{q}(t)\right\rangle\\
+&\left\langle\dot{q}(t)\phi_0'(t,0)\dot{q}(t)\phi_0'(t,0)\right\rangle\\
-&\left\langle\phi_0'(t,0)\dot{q}(t)\right\rangle^2-\left\langle\dot{q}(t)\phi_0'(t,0)\right\rangle^2\\
-&2\left\langle\phi_0'(t,0)\dot{q}(t)\right\rangle\left\langle\dot{q}(t)\phi_0'(t,0)\right\rangle\Big).
\end{split}
\end{equation}
We can use Wick's theorem to simplify the above equation. In the case we are considering, for example, the first term in the above equation can be expanded as
\begin{equation}
\begin{split}
&\left\langle\phi_0'(t,0)\dot{q}(t)\phi_0'(t,0)\dot{q}(t)\right\rangle\\
=&\left\langle\phi_0'(t,0)\dot{q}(t)\right\rangle\left\langle\phi_0'(t,0)\dot{q}(t)\right\rangle\\
+&\left\langle\phi_0'(t,0)\phi_0'(t,0)\right\rangle\left\langle\dot{q}(t)\dot{q}(t)\right\rangle\\
+&\left\langle\phi_0'(t,0)\dot{q}(t)\right\rangle\left\langle\dot{q}(t)\phi_0'(t,0)\right\rangle.
\end{split}
\end{equation}
One might note that the last two lines of \eqref{force fluctuation before expansion} can be deleted because of \eqref{phi'q'}. We keep them there because they can also be canceled exactly by Wick's expansion of the first four lines. After these cancellations we arrive at
\begin{equation}\label{force variance}
\begin{split}
\sigma_F(t)=\epsilon^2&\Big(\left\langle\phi_0'(t,0)^2\right\rangle\left\langle\dot{q}(t)^2\right\rangle\\
&+\left\langle\phi_0'(t,0)\dot{q}(t)\right\rangle\left\langle\dot{q}(t)\phi_0'(t,0)\right\rangle\Big).
\end{split}
\end{equation}
The second term of the above equation is just zero (see Eq.\eqref{phi'q'}). Also note that, in $1+1$ dimension, the term  $\left\langle\phi_0'^2\right\rangle=\frac{1}{2}\left(\langle\dot{\phi_0}^2\rangle+\left\langle\phi_0'^2\right\rangle\right)=\left\langle T_{00}\right\rangle=\left\langle T_{11}\right\rangle$, where $\left\langle T_{00}\right\rangle$ is the expectation value of vacuum energy density and $\left\langle T_{11}\right\rangle$ is the expectation value of vacuum stress. So, we obtain our final result for the fluctuation of force acting on the mirror:
\begin{equation}\label{fluctuation of force result}
\sigma_F(t)=\epsilon^2\left\langle\dot{q}(t)^2\right\rangle\left\langle T_{00}\right\rangle,
\end{equation}
which is proportional to the product of logarithmically divergent internal kinetic energy
\begin{equation}\label{effective mass}
\left\langle\dot{q}^2\right\rangle=\frac{\epsilon^2}{4\pi}\int_{-\infty}^{+\infty}dk\frac{|k|^3}{(k^2-\Omega^2)^2+\frac{\epsilon^4}{4}k^2}.
\end{equation}
and $k^2$ divergent vacuum energy density:
\begin{equation}\label{vacuum energy density}
\left\langle T_{00}\right\rangle=\frac{1}{4\pi}\int_{-\infty}^{+\infty}|k|dk=\infty.
\end{equation}
Here we see that the infinite value of the vacuum energy density does have physical significance. It enters the expression \eqref{fluctuation of force result} to characterize the fluctuation of the force acting on the mirror. Note that there is no gravitational interaction included in our mirror system. Therefore this is an example of a non-gravitational system where it is not the energy difference from the vacuum but the actual value of the vacuum energy that has physical significance.

As shown in \eqref{vacuum energy density}, the infinity appears in the value of the vacuum energy density is an ultraviolet divergence, i.e. it comes from the arbitrarily high frequency field modes. It is interesting that although the mirror is not sensitive to the high frequency field modes, the infinite value of the vacuum energy density still enters our expression \eqref{fluctuation of force result}. 

Infinite quantities are usually regarded as unphysical and some regularizations and renormalizations are needed. So it seems that the infinite value of the vacuum energy density $\left\langle T_{00}\right\rangle$ in \eqref{fluctuation of force result} does not make sense which is similar to what happened in the cosmological constant problem. However, it will be shown in the following sections that this infinite value does make sense because of two special properties of the quantum vacuum: the vacuum friction and the strong anti-correlation of vacuum fluctuations. In other words, the fluctuation of the force acting on the mirror at an instant of time is indeed infinite, but the mirror's position does not undergo infinite fluctuation. On the contrary, its fluctuating motion will be confined in a small region.

\section{the finite fluctuation of average force}\label{sec: force average}
Before allowing the mirror to start moving due to the fluctuating force acting on it, we would like to first calculate the fluctuation of the time average of the force acting on the static mirror. The first discussion of the average force fluctuation was given by Barton \cite{0305-4470-24-5-014,0305-4470-24-23-020}. The reasons to do this are (1) the force only determines the instantaneous acceleration of the mirror while the mirror's position is determined by the force integrated over time, i.e. it is determined by the time accumulation of the force. So we hope we can get some insight by first studying the fluctuation of the average force because the average is a kind of time accumulation; (2) any apparatus measuring the force cannot respond instantaneously. What the apparatus really measured is not the force at an instant of time but the average in a small time interval. If we finally get a finite result for the fluctuation of the average force, there is the possibility that the fluctuating motion of the mirror is finite. 

We will use the Gaussian function $\frac{1}{\sqrt{2\pi\sigma^2}}e^{-\frac{(t'-t)^2}{2\sigma^2}}$ to define the time average of the force as:
\begin{equation}\label{average force}
\bar{F}(t)=\frac{1}{\sqrt{2\pi\sigma^2}}\int_{-\infty}^{+\infty}F(t')e^{-\frac{(t'-t)^2}{2\sigma^2}}dt'.
\end{equation}
Its fluctuation is defined as
\begin{equation}
\sigma_{\bar{F}}(t)=\left\langle \bar{F}(t)^2\right\rangle-\left\langle \bar{F}(t)\right\rangle^2.
\end{equation}
Inserting \eqref{average force} into the above definition gives
\begin{equation}\label{fluctuation of force average}
\begin{split}
\sigma_{\bar{F}(t)}=\frac{1}{2\pi\sigma^2}&\int_{-\infty}^{+\infty}\int_{-\infty}^{+\infty}e^{-\frac{(t_1-t)^2+(t_2-t)^2}{2\sigma^2}}\\
&\cdot Corr(F(t_1),F(t_2))dt_1dt_2,
\end{split}
\end{equation}
where
\begin{equation}\label{correlation function}
\begin{split}
&Corr\left(F\left(t_1\right),F\left(t_2\right)\right)\\
=&\left[\left\langle F(t_1)F(t_2)\right\rangle-\left\langle F(t_1)\right\rangle\left\langle F(t_2)\right\rangle\right]
\end{split}
\end{equation}
is the correlation function between forces $F$ at time $t_1$ and $t_2$.
Next let us calculate the correlation function (\ref{correlation function}). Plugging \eqref{force} into the definition \eqref{correlation function} gives
\begin{equation}\begin{split}\label{correlation expansion}
&Corr(F(t_1),F(t_2))\\
=&\frac{\epsilon^2}{4}\bigg(\left\langle\phi_0'(t_1,0)\dot{q}(t_1)\phi_0'(t_2,0)\dot{q}(t_2)\right\rangle\\
+&\left\langle\phi_0'(t_1,0)\dot{q}(t_1)\dot{q}(t_2)\phi_0'(t_2,0)\right\rangle\\
+&\left\langle\dot{q}(t_1)\phi_0'(t_1,0)\phi_0'(t_2,0)\dot{q}(t_2)\right\rangle\\
+&\left\langle\dot{q}(t_1)\phi_0'(t_1,0)\dot{q}(t_2)\phi_0'(t_2,0)\right\rangle\\
-&\left\langle\phi_0'(t_1,0)\dot{q}(t_1)\right\rangle\left\langle\phi_0'(t_2,0)\dot{q}(t_2)\right\rangle\\
-&\left\langle\phi_0'(t_1,0)\dot{q}(t_1)\right\rangle\left\langle\dot{q}(t_2)\phi_0'(t_2,0)\right\rangle\\
-&\left\langle\dot{q}(t_1)\phi_0'(t_1,0)\right\rangle\left\langle\phi_0'(t_2,0)\dot{q}(t_2)\right\rangle\\
-&\left\langle\dot{q}(t_1)\phi_0'(t_1,0)\right\rangle\left\langle\dot{q}(t_2)\phi_0'(t_2,0)\right\rangle\bigg).
\end{split}
\end{equation}
Similar with the calculation of fluctuation of the force $\sigma_F$, we employ Wick's theorem to reduce the products of four operators to sum of products of pairs of operators to simplify the above equation. For example, the first term can be expanded as
\begin{equation}
\begin{split}
&\left\langle\phi_0'(t_1,0)\dot{q}(t_1)\phi_0'(t_2,0)\dot{q}(t_2)\right\rangle\\
=&\left\langle\phi_0'(t_1,0)\dot{q}(t_1)\right\rangle\left\langle\phi_0'(t_2,0)\dot{q}(t_2)\right\rangle\\
+&\left\langle\phi_0'(t_1,0)\phi_0'(t_2,0)\right\rangle\left\langle\dot{q}(t_1)\dot{q}(t_2)\right\rangle\\
+&\left\langle\phi_0'(t_1,0)\dot{q}(t_2)\right\rangle\left\langle\dot{q}(t_1)\phi_0'(t_2,0)\right\rangle.
\end{split}
\end{equation}
Applying Wick's theorem in (\ref{correlation expansion}) gives
\begin{equation}
\begin{split}
&Corr(F(t_1),F(t_2))\\
=&\epsilon^2\Big(\left\langle\phi_0'(t_1,0)\phi_0'(t_2,0)\right\rangle\left\langle\dot{q}(t_1)\dot{q}(t_2)\right\rangle\\
+&\left\langle\phi_0'(t_1,0)\dot{q}(t_2)\right\rangle\left\langle\dot{q}(t_1)\phi_0'(t_2,0)\right\rangle\Big).
\end{split}
\end{equation}
From (\ref{field expansion}) and (\ref{oscillator equation solution}) we can easily obtain
\begin{equation}
\left\langle\phi_0'(t_1,0)\phi_0'(t_2,0)\right\rangle=\frac{1}{4\pi}\int_{-\infty}^{+\infty}|k|e^{-i|k|(t_1-t_2)}dk,
\end{equation}
\begin{equation}
\begin{split}
&\left\langle\dot{q}(t_1)\dot{q}(t_2)\right\rangle\\
=&\frac{\epsilon^2}{4\pi}\int_{-\infty}^{+\infty}\frac{|k|^3}{(k^2-\Omega^2)^2+\frac{\epsilon^4}{4}k^2}e^{-i|k|(t_1-t_2)}dk,
\end{split}
\end{equation}
\begin{equation}
\begin{split}
&\left\langle\phi_0'(t_1,0)\dot{q}(t_2)\right\rangle\\
=&-i\frac{\epsilon}{4\pi}\int_{-\infty}^{+\infty}\frac{k|k|}{-k^2+\frac{i}{2}\epsilon^2|k|+\Omega^2}e^{-i|k|(t_1-t_2)}dk.
\end{split}
\end{equation}
Thus we reach an expression for the correlation function
\begin{equation}
\begin{split}\label{simmaf1f2}
&Corr(F(t_1),F(t_2))\\
=&\frac{\epsilon^4}{16\pi^2}\Bigg(\int_{-\infty}^{+\infty}|k|e^{-i|k|(t_1-t_2)}dk\\
&\cdot\int_{-\infty}^{+\infty}\frac{|k'|^3}{(k'^2-\Omega^2)^2+\frac{\epsilon^4}{4}k'^2}e^{-i|k'|(t_1-t_2)}dk'\\
+&\int_{-\infty}^{+\infty}\frac{k|k|}{-k^2+\frac{i}{2}\epsilon^2|k|+\Omega^2}e^{-i|k|(t_1-t_2)}dk\\
&\cdot\int_{-\infty}^{+\infty}\frac{k'|k'|}{-k'^2-\frac{i}{2}\epsilon^2|k'|+\Omega^2}e^{-i|k'|(t_1-t_2)}dk'\Bigg).
\end{split}
\end{equation}
Plugging (\ref{simmaf1f2}) into (\ref{fluctuation of force average}) and changing the order of integration gives
\begin{equation}\begin{split}
&\sigma_{\bar{F}(t)}=\frac{\epsilon^4}{16\pi^2}\cdot\frac{1}{2\pi\sigma^2}\\
\times&\int_{-\infty}^{+\infty}\int_{-\infty}^{+\infty}|k|\cdot\frac{|k'|^3}{(k'^2-\Omega^2)^2+\frac{\epsilon^4}{4}k'^2}dkdk'\\
\times&\int_{-\infty}^{+\infty}e^{-\frac{(t_1-t)^2}{2\sigma^2}-i(|k|+|k'|)t_1}dt_1\\
\times&\int_{-\infty}^{+\infty}e^{-\frac{(t_2-t)^2}{2\sigma^2}+i(|k|+|k'|)t_2}dt_2\\
=&\frac{\epsilon^4}{16\pi^2}\int_{-\infty}^{+\infty}\int_{-\infty}^{+\infty}\frac{|k||k'|^3}{(k'^2-\Omega^2)^2+\frac{\epsilon^4}{4}k'^2}\\
&\cdot e^{-\sigma^2(|k|+|k'|)^2}dkdk'\\
\leq &\frac{\epsilon^4}{16\pi^2}\int_{-\infty}^{+\infty}|k|e^{-\sigma^2 k^2}dk\\
&\cdot\int_{-\infty}^{+\infty}\frac{|k'|^3}{(k'^2-\Omega^2)^2+\frac{\epsilon^4}{4}k'^2}e^{-\sigma^2 k'^2}dk'\\
<&+\infty.
\end{split}
\end{equation}
Thus we get a finite result for the fluctuation of the time-averaged force. The finiteness of the fluctuation of the force average is closely related to the strong anti-correlation property of the vacuum fluctuations. Detailed analysis of this property will be given in section \ref{sec: discussions}.

\section{the vacuum friction: damping force when the mirror starts to move}\label{sec:Motion effects of the mirror}
In this section we allow the mirror to begin to move. We are interested in the question of how the mirror move if we release it at time $t=0$. One might naively think that the mirror's position will fluctuate infinitely under the infinite fluctuating force, although such a result must be unphysical. However, as we stated in the beginning of the last section, the force can only determine the instantaneous acceleration of the mirror, while the position of the mirror is determined by the time integration of the force. We see from the last section that the fluctuation of average force is finite. This gives a hope that the fluctuation of the position of the mirror, which is driven by the force exerted on it, might be finite. To calculate this position fluctuation, i.e. the mean-squared displacement, we need to figure out the equation of motion of the mirror.

P.C.W.Davies has suggested that the quantum vacuum may in certain circumstances be regarded as a type of fluid medium exhibiting friction \cite{1464-4266-7-3-006}. We expect that when our mirror starts to move, it will experience a frictional force damping its motion. This force is important in constructing the equation of motion. This section we will give the detailed analysis about this force and the construction of the equation of motion will be given in the next section. 

Unlike in the previous sections where we held the mirror fixed at location $x=0$, in this section we specify the mirror move along a generic trajectory and investigate the damping force acting on it by the field.

Now let us calculate the damping force in detail. Consider the mirror is moving along a generic trajectory $x=X\left(t\left(\tau\right)\right)$, where $\tau$ is the proper time associated with this trajectory:
\begin{equation} 
t(\tau)=\int_0^{\tau}\gamma(t(\tau')) d\tau',
\end{equation}
where $\gamma(t)=\frac{1}{\sqrt{1-\dot{X}^2}}$ is the Lorentz factor and $\dot{}$ denote derivative with respect to coordinate time $t$ as before, i.e. $\dot{X}(t)=\frac{dX(t)}{dt}$. The action of the moving mirror is 
\begin{equation}\label{moving action}
\begin{split}
S&=\frac{1}{2}\iint\left(\left(\frac{\partial\phi}{\partial t}\right)^2-\left(\frac{\partial\phi}{\partial x}\right)^2\right)dt dx\\
&+\frac{1}{2}\int\left(\left(\frac{d q}{d\tau}\right)^2-\Omega^2q^2\right)d\tau\\
&+\epsilon\int\frac{d\phi}{d\tau}\left(t\left(\tau\right),X(t\left(\tau\right)\right))q\left(t\left(\tau\right)\right)d\tau.
\end{split}
\end{equation}
Note that $X$ and $q$ are different things. $X$ is the mirror's position which is moving ``in space'' while $q$ is the mirror's ``internal'' degree of freedom which is \emph{Not} oscillating ``in space''.

The equations of motion for the field $\phi$ and the internal harmonic oscillator $q$ now become
\begin{equation}\label{motion phi equation}
\ddot{\phi}-\phi''=-\epsilon\dot{q}\delta\left(x-X\left(t\right)\right),
\end{equation}
\begin{equation}\label{motion q equation}
\frac{d^2 q}{d\tau^2}+\Omega^2 q^2=\epsilon\frac{d\phi}{d\tau}(t(\tau),X(t(\tau))).
\end{equation}
Similar to the static mirror case, the solution of \eqref{motion phi equation} is of the following form
\begin{equation}\label{motion phi solution}
\phi(t,x)=\phi_0(t,x)-\frac{\epsilon}{2}q(t'),
\end{equation}
where the retarded time $t'$ is determined by the following equation
\begin{equation}\label{line eq}
t-t'=|x-X(t')|.
\end{equation}
Substituting \eqref{motion phi solution} into the equation of motion for the internal harmonic oscillator \eqref{motion q equation} gives
\begin{equation}\label{motion driven sho equation}
\frac{d^2 q}{d\tau^2}+\frac{\epsilon^2}{2}\frac{d q}{d\tau}+\Omega^2 q=\epsilon\frac{d\phi_0}{d\tau}(t(\tau),X(t(\tau))).
\end{equation}
Similar to solution \eqref{static oscillator equation solution} for the static mirror, the solution of the above equation of motion \eqref{motion driven sho equation} of the internal driven damped harmonic oscillator  is
\begin{equation}\label{motion q solution}
\begin{split}
q(t(\tau))=\frac{1}{\omega}\int_{-\infty}^{\tau}&e^{-a(\tau-\tau')}\sin(\omega(\tau-\tau'))\\
&\cdot\epsilon\frac{d\phi_0}{d\tau'}(t(\tau'),X(t(\tau')))d\tau',
\end{split}
\end{equation}
where $a=\frac{\epsilon^2}{4}$ and $\omega=\sqrt{\Omega^2-\frac{\epsilon^4}{16}}$ are the same with those in \eqref{static oscillator equation solution}. One key difference of the solution \eqref{motion q solution} from the static case \eqref{static oscillator equation solution} is that when the mirror moves, the driving force changes, which could result in the deviation of the $q$'s motion from its steady oscillation state \eqref{oscillator equation solution} or \eqref{internal degree expansion}.

\begin{figure}
\centering
\includegraphics[scale=0.5]{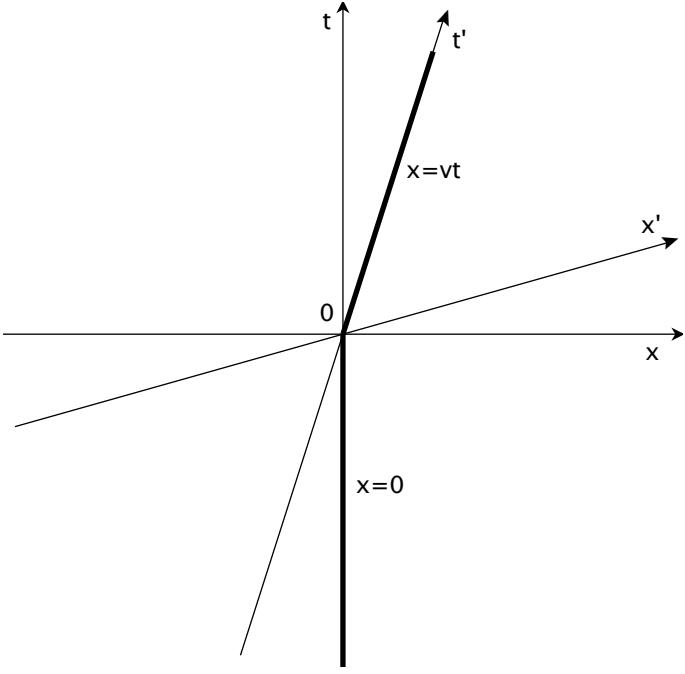}
\caption{\label{jump}The trajectory for a mirror who initially stays at rest and then jumps to move with a constant velocity $v$ at time $t=0$.}
\end{figure}

Here we consider the force in the mirror's instantaneous rest frame. In this frame, the force acting on each side of the mirror by the field is the form of $T_{\mu\nu}x^{\mu}x^{\nu}$, where $x^{\mu}=\gamma(\dot{X},1)$ is a unit spacelike vector which is orthogonal to the four velocity of the mirror. Thus the force in the moving mirror's instantaneous rest frame is defined as
\begin{equation}\begin{split}\label{motion force}
F(t)=&\lim_{x\to 0^+}\left(T_{\mu\nu}(x_-)x^{\mu}x^{\nu}-T_{\mu\nu}(x_+)x^{\mu}x^{\nu}\right)\\
=\gamma^2\lim_{x\to 0^+}\Big(&(T_{00}(x_-)-T_{00}(x_+))\dot{X}^2\\
+&2(T_{01}(x_-)-T_{01}(x_+))\dot{X}\\
+&(T_{11}(x_-)-T_{11}(x_+))\Big),
\end{split}\end{equation}
where $x_-=(t,X(t)-x)$ and $x_+=(t,X(t)+x)$ ($x\geqslant0$) are two spacetime points which are symmetrically located on the two sides of the mirror. $T_{00}$, $T_{01}$ and $T_{11}$ are components of stress-energy tensor of type $(0,2)$ which in $(t,x)$ coordinates are defined as
\begin{eqnarray}
T_{00}&=&\frac{1}{2}\left(\dot{\phi}^2(t,x)+\phi'^2(t,x)\right),\label{t00}\\
T_{01}&=&\frac{1}{2}\left(\dot{\phi}(t,x)\phi'(t,x)+\phi'(t,x)\dot{\phi}(t,x)\right),\label{t01}\\
T_{11}&=&\frac{1}{2}\left(\dot{\phi}^2(t,x)+\phi'^2(t,x)\right)\label{t11}.
\end{eqnarray}
From \eqref{motion phi solution} and \eqref{line eq} we can get the time and space derivatives of the field $\phi$:
\begin{equation}\label{moving time derivative}
\dot{\phi}(t,x) = \left\{
  \begin{array}{l l l}
    \dot{\phi_0}(t,x)-\frac{\epsilon}{2}\left(\frac{\dot{q}(t')}{1+\dot{X}(t')}\right), & \text{if $x<X(t')$}\\
    \dot{\phi_0}(t,x)-\frac{\epsilon}{2}\left(\frac{\dot{q}(t')}{1-\dot{X}(t')}\right), & \text{if $x>X(t')$}\\
  \end{array} \right.
\end{equation}
\begin{equation}\label{moving space derivative}
\phi'(t,x) = \left\{
\begin{array}{l l l}
    \phi_0'(t,x)-\frac{\epsilon}{2}\left(\frac{\dot{q}(t')}{1+\dot{X}(t')}\right), & \text{if $x<X(t')$}\\
    \phi_0'(t,x)+\frac{\epsilon}{2}\left(\frac{\dot{q}(t')}{1-\dot{X}(t')}\right), & \text{if $x>X(t')$}\\
\end{array} \right.
\end{equation}
Similar to the static mirror case, we can first insert \eqref{moving time derivative} and \eqref{moving space derivative} into \eqref{t00}, \eqref{t01} and \eqref{t11} to express the stress-energy tensor components in terms of $\phi_0$ and $q$, then plug these expressions into \eqref{motion force} to get the force. The result is
\begin{equation}\label{motion force expression}
F(t)=-\epsilon\gamma^2\{\dot{q}(t)\left(\phi_0'(t,X(t))+\dot{X}\dot{\phi_0}(t,X(t))\right)\},
\end{equation}
where the curly bracket $\{\}$ represents the symmetrization operation \eqref{symmetric operation} as before. This formula can be understood as following: remember that the force we are calculating is evaluated in the mirror's instantaneous rest frame, which should have the same form with the static mirror case \eqref{force} when expressed in terms of its own instantaneous rest frame coordinates $(t',x')$ (see Fig. \ref{jump}):
\begin{equation}\label{rest frame force}
F(t')=-\epsilon\left\{\frac{\partial\phi_0(t',0)}{\partial x'}\cdot\frac{dq(t')}{dt'}\right\}.
\end{equation}
Changing the above expression \eqref{rest frame force} from $(t',x')$ coordinate system to the laboratory coordinate system $(t,x)$ leads to exactly the force expression \eqref{motion force expression}, which is expressed in terms of laboratory frame coordinates. 

Next we can insert the expression \eqref{field expansion} for the incident wave $\phi_0$  and the expression \eqref{motion q solution} for the internal degree of freedom $q$ into \eqref{motion force expression} to get the mean motional force exerted on the mirror by the field when the mirror is moving along a generic trajectory $x=X(t(\tau))$. The result is:
\begin{equation}\label{moving force}\begin{split}
&\left\langle F(t(\tau))\right\rangle \\
=&-\frac{1}{2}\frac{\epsilon^2}{4\pi}\gamma(t(\tau))\frac{1}{\omega}\int_{-\infty}^{+\infty}dk\int_{-\infty}^{\tau}d\tau'\gamma(t(\tau'))\\
\times&\Big[-k\left(1+\dot{X}(t(\tau))\dot{X}(t(\tau'))\right)\\
&+|k|\left(\dot{X}(t(\tau))+\dot{X}(t(\tau'))\right)\Big]\\
\times&\Big[-a\sin(\omega(\tau-\tau'))+\omega\cos(\omega(\tau-\tau'))\Big]\\
\times&\exp\Bigg(-a(\tau-\tau')+i\Big[|k|\int_{\tau'}^{\tau}\gamma(t(\tau''))d\tau''\\
&\quad\quad\quad-k(X(t(\tau))-X(t(\tau')))\Big]\Bigg)\\
+&c.c.
\end{split}
\end{equation}

\begin{figure}
\centering
\includegraphics[scale=0.5]{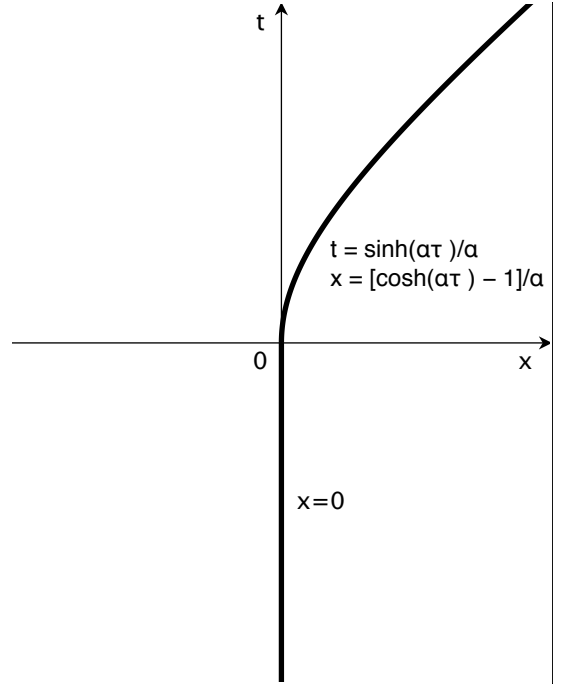}
\caption{\label{mirror trajectory}The trajectory for a mirror who initially stays at rest and then starts to move with constant acceleration $\alpha$ at $t=0$.}
\end{figure}

The above expression for the force is quite complicated. However, remember that what we are interested in is the damping force when the mirror starts to move due to the quantum fluctuations of the field after we release it. So let us consider a motion in which the mirror initially stays at the origin for a long time and then starts to move with constant acceleration $\alpha$ along the following trajectory when $t\geq 0$ (As shown in FIG. \ref{mirror trajectory}):
\begin{equation}
\left\{
\begin{array}{l l}
t=\frac{1}{\alpha}\sinh(\alpha\tau),\\
x=\frac{1}{\alpha}[\cosh(\alpha\tau)-1],
\end{array}\right.
\end{equation}
where $\tau$ is the proper time of the trajectory as before. Plugging this trajectory into \eqref{moving force}, we obtain that, when the velocity changes from $0$ to $v\approx \alpha\tau\ll 1$, the mean motional force is
\begin{equation}\label{damping formula}
\left\langle F\right\rangle=\left[-\frac{\epsilon^4}{4\pi}\int_0^{+\infty}\frac{k^3dk}{(k^2-\Omega^2)^2+\frac{\epsilon^4}{4}k^2}\right]v.
\end{equation}
The above formula shows that the quantum vacuum does serve as a fluid medium in the sense that our mirror, if initially stays at rest, would experience a friction force when it starts to move . 

Here we emphasize that the $v$ in the above formula \eqref{damping formula} should be understood as the \emph{velocity changes relative to the mirror's original instantaneous rest frame before the acceleration happened where the internal harmonic oscillator has already reached a steady oscillation state}. For the constantly moving mirror trajectory $X(t)\equiv vt$, the formula \eqref{damping formula} dose not apply and the general motional force expression \eqref{moving force} gives zero force. This zero force is just the requirement of Lorentz invariance. 

In fact, the friction force \eqref{damping formula} arises from the Doppler shift of the vacuum modes due to the changing velocity of the mirror. Referring back to FIG. \ref{mirror energy transfer process}, the mirror absorbs both the left moving field modes (wave number $k<0$) and the right moving field modes (wave number $k>0$). The rate at which the mirror absorbs energy from a mode depends on the internal velocity of the internal oscillator $\dot{q}$. Because of the linearity of the system and the lack of any correlations between these modes, the only component of the oscillator motion that is important is the motion in $q$ induced by that same mode earlier in time. If the mirror starts to move, the oscillator saw that mode at an earlier time with a frequency that was Doppler shifted from what it sees now - modes coming from one side are red shifted and the other blue shifted. Thus different amounts of energy (and thus of momentum) will be absorbed from the two directions. The emitted momenta, on the other hand, is always balanced between the two sides, and supplies no force to the mirror (see top line in FIG. \ref{mirror energy transfer process}). 

For example, if the mirror moves to the right with velocity $v$, the right moving modes with frequency $|k|$ will be red shifted to frequency $\left(\frac{1-v}{1+v}\right)^{1/2}|k|$ and the left moving modes with the same frequency will be blue shifted to frequency $\left(\frac{1+v}{1-v}\right)^{1/2}|k|$. So the symmetry between the left moving modes and right moving modes is broken. This asymmetry will result in the force imbalance since the mirror will absorb more momentum from left moving modes than from right, which gives a net force to the left to resist the mirror's motion. To understand this in detail, let us investigate the following trajectory (As shown in FIG. \ref{jump}):
\begin{equation}\label{instant jump trajectory}
X(t)=\left\{
\begin{array}{l l}
0, &  \quad\text{if $t<0$}\\
vt, & \quad\text{if $t\geq 0$}
\end{array}\right.
\end{equation}
For this trajectory the mirror is initially static until it starts to move with constant velocity at time $t=0$. Direct calculation using the general mean motional force formula \eqref{moving force} shows that when $t\geq 0$, the friction force is
\begin{equation}\label{jumping force general}
\begin{split}
&\left\langle F(t(\tau))\right\rangle\\
=&-\frac{1}{2}\frac{\epsilon^2}{4\pi}\frac{1}{\omega}e^{-a\tau}\int_{0}^{+\infty} k\Big[\left(\frac{1+v}{1-v}\right)^{\frac{1}{2}}e^{ik\left(\frac{1+v}{1-v}\right)^{\frac{1}{2}}\tau}\\
&\quad\quad\quad-(\frac{1-v}{1+v})^{\frac{1}{2}}e^{ik(\frac{1-v}{1+v})^{\frac{1}{2}}\tau}\Big]\\
\cdot&\frac{-ik\omega\cos(\omega\tau)-(\Omega^2-\frac{i}{4}\epsilon^2k)\sin(\omega\tau)}{\Omega^2-k^2-\frac{i}{2}\epsilon^2k}dk+c.c.
\end{split}\end{equation}
Note that the exponential factor $e^{-a\tau}$ appearing in the above expression implies that after a long time the force would decrease to zero. This is just the requirement of Lorentz invariance, since after long times the mirror's memory would fade away and it would not remember what it did long before and thus can be regarded as a moving mirror with constant velocity $v$, which should experience zero friction force.

What's interesting is the force at the time $t=0$, it is
\begin{equation}\label{jumping force}
\begin{split}
\left\langle F(0)\right\rangle=&-\frac{\epsilon^4}{8\pi}\left[\left(\frac{1+v}{1-v}\right)^{\frac{1}{2}}-\left(\frac{1-v}{1+v}\right)^{\frac{1}{2}}\right]\\
&\cdot\int_0^{+\infty}\frac{k^3dk}{(\Omega^2-k^2)^2+\frac{1}{4}\epsilon^4k^2},
\end{split}
\end{equation}
which is in agreement with our previous result \eqref{damping formula} at small velocity approximation. Here the two factors $(\frac{1+v}{1-v})^{\frac{1}{2}}$ and $(\frac{1-v}{1+v})^{\frac{1}{2}}$ are exactly relativistic Doppler shift factors for an observer moving toward or away from a light source with velocity $v$. 

Next we reproduce the above result \eqref{jumping force} by a different method to reveal the role played by Doppler effect. We will consider everything in the mirror's instantaneous rest frame.

For the trajectory \eqref{instant jump trajectory}, when $t<0$, the mirror's rest frame is $(t,x)$ coordinate system and the field $\phi_0$ is expanded as the sum of positive frequency modes $\frac{e^{-i(|k|t-kx)}}{\sqrt{4\pi|k|}}$ with coefficients $a_k$ and negative frequency modes $\frac{e^{+i(|k|t-kx)}}{\sqrt{4\pi|k|}}$ with coefficients $a_k^{\dag}$ (see \eqref{field expansion}). When $t\geq 0$, the mirror's rest frame is $(t',x')$ coordinate system (see FIG (\ref{jump})) and the same field $\phi_0$ is expanded as:
\begin{equation}\begin{split}
\phi_0(t',x')=\int_{-\infty}^{+\infty}\frac{dk'}{\sqrt{4\pi|k'|}}&\Big(b_{k'}e^{-i\left(|k'|t'-k'x'\right)}\\
+&b_{k'}^{\dag}e^{i\left(|k'|t'-k'x'\right)}\Big),
\end{split}\end{equation}
where the wave numbers $k'$ in $(t',x')$ coordinate system are Doppler shifted from the wave numbers $k$ in $(t,x)$ system to:
\begin{equation}
k'=\left\{
\begin{array}{l l}
(\frac{1+v}{1-v})^{1/2}k, \quad when \quad k<0\\
(\frac{1-v}{1+v})^{1/2}k, \quad when \quad k>0,
\end{array}\right.
\end{equation}
and correspondingly, the operator coefficients $b_{k'}$ and $a_k$ are related by
\begin{equation}
b_{k'}=\left\{
\begin{array}{l l}
(\frac{1-v}{1+v})^{1/4}a_k, \quad when \quad k<0\\
(\frac{1+v}{1-v})^{1/4}a_k, \quad when \quad k>0.
\end{array}\right.
\end{equation}
Now expand the solution of the internal harmonic oscillator \eqref{motion q solution} in terms of the new operators $b_{k'}$ and $b_{k'}^{\dag}$. For the trajectory \eqref{instant jump trajectory} we are considering, the oscillation pattern right before $\tau=0$ is
\begin{equation}
q(\tau)=q^L(\tau)+q^R(\tau),
\end{equation}
where
\begin{equation}\begin{split}
q^L(\tau)&=-i\epsilon\int_{-\infty}^{0}dk'(\frac{1-v}{1+v})^{1/2}\sqrt{\frac{|k'|}{4\pi}}\\
&\cdot\Bigg(\frac{b_{k'}e^{-i(\frac{1-v}{1+v})^{1/2}|k'|\tau}}{-(\frac{1-v}{1+v})k'^2-\frac{i}{2}\epsilon^2(\frac{1-v}{1+v})^{1/2}|k'|+\Omega^2}\\
&-\frac{b_{k'}^{\dag}e^{i(\frac{1-v}{1+v})^{1/2}|k'|\tau}}{-(\frac{1-v}{1+v})k'^2+\frac{i}{2}\epsilon^2(\frac{1-v}{1+v})^{1/2}|k'|+\Omega^2}\Bigg)
\end{split}\end{equation}
and
\begin{equation}\begin{split}
q^R(\tau)&=-i\epsilon\int_{0}^{+\infty}dk'(\frac{1+v}{1-v})^{1/2}\sqrt{\frac{|k'|}{4\pi}}\\
&\cdot\Bigg(\frac{b_{k'}e^{-i(\frac{1+v}{1-v})^{1/2}|k'|\tau}}{-(\frac{1+v}{1-v})k'^2-\frac{i}{2}\epsilon^2(\frac{1+v}{1-v})^{1/2}|k'|+\Omega^2}\\
&-\frac{b_{k'}^{\dag}e^{i(\frac{1+v}{1-v})^{1/2}|k'|\tau}}{-(\frac{1+v}{1-v})k'^2+\frac{i}{2}\epsilon^2(\frac{1+v}{1-v})^{1/2}|k'|+\Omega^2}\Bigg).
\end{split}\end{equation}
Unlike the oscillation \eqref{oscillator equation solution}, after the mirror did an instant jump in velocity at $\tau=0$, the mirror sees that the field modes are Doppler shifted and the oscillation of $q$ above is no longer steady state relative to the field in the new $(t',x')$ frame. Driven by these Doppler shifted modes, the oscillation will change and eventually settle down to a new steady state, with the same frequencies as the driving field modes. Once it reached the final steady state again, the frictional force would again become zero as predicted in \eqref{jumping force general}. However, during this process, there will be imbalance between the absorbed momentum from the left and the right, which gives the non-zero friction force. In fact, in the mirror's frame, the force \eqref{motion force expression} would reduce to \eqref{force} or \eqref{momentum force}. And the average momentum absorbed per unit time from right and from left at the jumping point $\tau=0$ are
\begin{equation}\label{left force}
\begin{split}
&\left\langle\left\{\epsilon\frac{d{\phi}_0^L}{d\tau}\frac{dq^L}{d\tau}\right\}\right\rangle\\
=&\frac{\epsilon^4}{8\pi}(\frac{1+v}{1-v})^{\frac{1}{2}}\int_0^{+\infty}\frac{k^3dk}{(\Omega^2-k^2)^2+\frac{1}{4}\epsilon^4k^2},
\end{split}
\end{equation}
and
\begin{equation}\label{right force}
\begin{split}
&\left\langle\left\{\epsilon\frac{d{\phi}_0^R}{d\tau}\frac{dq^R}{d\tau}\right\}\right\rangle\\
=&\frac{\epsilon^4}{8\pi}(\frac{1-v}{1+v})^{\frac{1}{2}}\int_0^{+\infty}\frac{k^3dk}{(\Omega^2-k^2)^2+\frac{1}{4}\epsilon^4k^2},
\end{split}
\end{equation}
where the curly brackets $\{\}$ denote the symmetrization operation as defined in \eqref{symmetric operation} and we have dropped the terms $\left\langle \epsilon\frac{d{\phi}_0^L}{d\tau}\frac{dq^R}{d\tau}\right\rangle$ and $\left\langle \epsilon\frac{d{\phi}_0^R}{d\tau}\frac{dq^L}{d\tau}\right\rangle$ since they are zero. Note that the force \eqref{jumping force} is exactly the difference of \eqref{left force} and \eqref{right force}. Thus we can come to the conclusion that if the mirror starts to move from zero velocity and has already acquired a velocity, for example, to the right, it would absorb more momentum from the right per unit time than from the left. The difference is determined by the Doppler shift factors. It is this imbalance in absorbed momentum from different directions that leads to the non-zero frictional force.

An important lesson we learn from the above analyses is that the non-zero frictional force happens only when the oscillation of the internal harmonic oscillator has deviated from the steady state. That's why we emphasized after the formula \eqref{damping formula} that the $v$ should be understood as the \emph{velocity changes relative to the mirror's original instantaneous rest frame before the acceleration happened where the internal harmonic oscillator has already reached a steady oscillation state}.

One might also worry about the logarithmically divergent proportional constant in the force expression \eqref{damping formula} for the damping force. It does not matter because the damping force is not an observable quantity. The physically observable quantity is the motion of the mirror which depends on the time average of the force, which is proved to be finite in the last section, or the movement of the mirror under the influence of this divergent force. It turns out in the following sections that the effective mass of the mirror is also logarithmically divergent, which exactly cancels the divergence of the damping force to give a finite value of damping ratio.

\section{the mirror's equation of motion}\label{sec: mirror equation of motion}
Unlike in the last section, where we specified the trajectory the mirror moved along, in this section we release the mirror and let it move freely under the fluctuating force exerting on it by the field. To do this, we add an extra term (the first one) in the action \eqref{moving action} such that:
\begin{equation}\label{released moving action}
\begin{split}
S=&-M\int d\tau\\
&+\frac{1}{2}\iint\left(\left(\frac{\partial\phi}{\partial t}\right)^2-\left(\frac{\partial\phi}{\partial x}\right)^2\right)dt dx\\
&+\frac{1}{2}\int\left(\left(\frac{d q}{d\tau}\right)^2-\Omega^2q^2\right)d\tau\\
&+\epsilon\int\frac{d\phi}{d\tau}\left(t\left(\tau\right),X(t\left(\tau\right)\right))q\left(t\left(\tau\right)\right)d\tau,
\end{split}
\end{equation}
where $M$ is the mirror's bare mass. One can derive the mirror's equation of motion directly from this action (see Appendix \ref{derive action motion}). However, to express the equation of motion in terms of the force in the mirror's instantaneous rest frame we derived in the last section, we choose another way--first derive the stress-energy tensor of the whole system and apply the continuity equation $\partial_{\nu}T^{\mu\nu}=0$ to obtain the equation of motion.

The stress-energy tensor of the whole sytem is (see \eqref{mirror stress-energy tensor} in Appendix \ref{derive stress-energy tensor})
\begin{eqnarray}
T^{00}&=&\frac{1}{2}\left(\dot{\phi}^2+\phi'^2\right)+\gamma M_{eff}\delta\left(x-X(t)\right),\\
T^{01}&=&T^{10}=-\left\{\dot{\phi}\phi'\right\}+\gamma M_{eff}\dot{X}\delta\left(x-X(t)\right),\\
T^{11}&=&\frac{1}{2}\left(\dot{\phi}^2+\phi'^2\right)+\gamma M_{eff}\dot{X}^2\delta\left(x-X(t)\right),
\end{eqnarray}
where the effective mass includes the mirror's bare mass $M$ and the energy of the internal harmonic oscillator (see \eqref{effective mass expression} in Appendix \ref{derive stress-energy tensor}):
\begin{equation}
M_{eff}=M+\frac{1}{2}\left(\frac{dq}{d\tau}\right)^2+\frac{1}{2}\Omega^2q^2.
\end{equation}
Next we apply the continuity equation $\partial_{\nu}T^{\mu\nu}=0$ to the above stress energy tensor. Also using the equation of motion of the field \eqref{motion phi equation}, we obtain the equation of energy conservation (for the case $\mu=0$):
\begin{equation}\label{energy conservation equation of motion}
\frac{d}{dt}\left(\gamma M_{eff}\right)=\epsilon\left\{\dot{q}(t)\dot{\phi}(t,X(t))\right\},
\end{equation}
and the equation of momentum conservation (for the case $\mu=1$):
\begin{equation}\label{equation of motion in lab frame}
\frac{d}{dt}\left(\gamma M_{eff}\dot{X}\right)=-\epsilon\left\{\dot{q}(t)\phi'(t,X(t))\right\}.
\end{equation}
Note that the term $-\epsilon\left\{\dot{q}(t)\phi'(t,X(t))\right\}$ in the above equation \eqref{equation of motion in lab frame} represents the force exerted on the mirror in laboratory frame, which is different from the force in the mirror's instantaneous rest frame \eqref{motion force expression}. We can derive the equation of motion in the mirror's instantaneous rest frame from \eqref{energy conservation equation of motion} and \eqref{equation of motion in lab frame} by performing a Lorentz boost and thus prove that \eqref{motion force expression} does serve as the force in the mirror's instantaneous rest frame. First, we let
\begin{equation}\label{momentum relation}
\begin{pmatrix}
E\\
P
\end{pmatrix}
=
\begin{pmatrix}
\gamma M_{eff}\\
\gamma M_{eff}\dot{X}
\end{pmatrix}
\end{equation}
be the energy-momentum vector of the mirror in the laboratory frame $(t,x)$ (as shown in FIG. \ref{jump}). Assuming that at some moment the mirror's instantaneous rest frame is $(t',x')$ (as shown in FIG. \ref{jump}), i.e. the frame $(t',x')$ is moving with velocity $\dot{X}$ with respect to the frame $(t,x)$. Then the energy-momentum vector $\begin{pmatrix}
E'\\
P'
\end{pmatrix}$ in $(t',x')$ frame is related to the energy-momentum vector $\begin{pmatrix}
E\\
P
\end{pmatrix}$ in $(t,x)$ frame by the Lorentz boost:
\begin{equation}\label{Lorentz boost momentum}
\begin{pmatrix}
E'\\
P'
\end{pmatrix}
=
\begin{pmatrix}
\gamma & -\gamma\dot{X}\\
-\gamma\dot{X} & \gamma
\end{pmatrix}
\begin{pmatrix}
E\\
P
\end{pmatrix}
\end{equation}
Differentiating $P'$ in \eqref{Lorentz boost momentum} with respect to the mirror's proper time $t'$ and using \eqref{energy conservation equation of motion}, \eqref{equation of motion in lab frame} and \eqref{momentum relation} yields:
\begin{equation}\label{phi mirror equation of motion}
\begin{split}
\frac{dP'}{dt'}&=\gamma\frac{dP'}{dt}\\
&=\gamma\left(-\gamma\dot{X}\frac{dE}{dt}+\gamma\frac{dP}{dt}\right)\\
&=-\epsilon\gamma^2\left\{\dot{q}(t)\left(\phi'(t,X(t))+\dot{X}\dot{\phi}(t,X(t))\right)\right\}.
\end{split}
\end{equation}

Note that in usual mathematical sense the field $\phi$ is not differentiable on the mirror's path $(t,X(t))$ since $\dot{\phi}$ and $\phi'$ have jump discontinuities there (see \eqref{moving time derivative} and \eqref{moving space derivative}). For this type of discontinuity, the values of $\dot{\phi}$ and $\phi'$ at $(t,X(t))$ are not defined and may have any value. However, it is natural to define the derivative as the average of left derivative and right derivative, i.e.
\begin{eqnarray}
\dot{\phi}(t,X(t))=&\lim_{x\to 0^+}\frac{\dot{\phi}(t,X(t)+x)+\dot{\phi}(t,X(t)-x)}{2},\\
\phi'(t,X(t))=&\lim_{x\to 0^+}\frac{\phi'(t,X(t)+x)+\phi'(t,X(t)-x)}{2}.
\end{eqnarray}
Applying the above definition to \eqref{moving time derivative} and \eqref{moving space derivative} yields
\begin{equation}\label{phi0 substitute phi}
\begin{split}
&\phi'(t,X(t))+\dot{X}\dot{\phi}(t,X(t))\\
=&\phi_0'(t,X(t))+\dot{X}\dot{\phi_0}(t,X(t))
\end{split}
\end{equation}
Thus we can replace the $\phi$ in \eqref{phi mirror equation of motion} by $\phi_0$ to obtain the mirror's equation of motion in its instantaneous rest frame:
\begin{equation}
\frac{dP'}{dt'}=-\epsilon\gamma^2\left\{\dot{q}(t)\left(\phi_0'(t,X(t))+\dot{X}\dot{\phi_0}(t,X(t))\right)\right\}.
\end{equation}
Note that the right hand side of the above equation exactly agree with the force expression \eqref{motion force expression} that we derived in the last section by a different method.

To analyze the fluctuating motion of the mirror we next express the mirror's equation of motion in the laboratory frame in terms of the force in its instantaneous rest frame by simple manipulations of the energy-momentum conservation equations \eqref{energy conservation equation of motion} and \eqref{equation of motion in lab frame}:
\begin{equation}
\begin{split}
&\gamma M_{eff}\frac{d^2X}{dt^2}\\
=&-\dot{X}\frac{d}{dt}\left(\gamma M_{eff}\right)-\epsilon\left\{\dot{q}(t)\phi'(t,X(t))\right\}\\
=&-\epsilon\left\{\dot{q}(t)\left(\phi'(t,X(t))+\dot{X}\dot{\phi}(t,X(t))\right)\right\}\\
=&\frac{1}{\gamma^2}\left(-\epsilon\gamma^2\left\{\dot{q}(t)\left(\phi_0'(t,X(t))+\dot{X}\dot{\phi_0}(t,X(t))\right)\right\}\right)
\end{split}
\end{equation}
where we have used \eqref{phi0 substitute phi} to replace $\phi$ by $\phi_0$ in the last line of the above equation. Note that the expression inside the parentheses of the last line is exactly the force $F$ \eqref{motion force expression} in the mirror's instantaneous rest frame that we derived in the last section, thus we reach the following equation of motion which relates the mirror's acceleration with the force $F$ in the mirror's instantaneous rest frame:
\begin{equation}\label{final lab frame equation of motion}
\gamma^3 M_{eff}\frac{d^2X}{dt^2}=F.
\end{equation}
We will analyze the fluctuating motion of the mirror using the above equation \eqref{final lab frame equation of motion}.

\section{confined fluctuating motion of the mirror}\label{sec: confined motion}
The situation we are considering is that we first hold the mirror fixed for a long time and then release it at time $t=0$. The mirror's position will then start to fluctuate due the quantum fluctuating force acting on it. We assume that the time scale of the period of the fluctuating motion is small enough that the oscillations of the internal harmonic oscillator would approximately stay in the steady state relative to the laboratory frame. For simplicity, we also assume that the velocity of the mirror is small, then we can neglect the $\gamma^3$ term in \eqref{final lab frame equation of motion} and the equation of motion becomes
\begin{equation}\label{newton eq}
M_{eff}\frac{d^2X}{dt^2}=F,
\end{equation}
We can rewrite the equation of motion \eqref{newton eq} as
\begin{equation}\label{newton eq revised}
\frac{d^2X}{dt^2}-\frac{\left\langle F\right\rangle}{M_{eff}}=\frac{F-\left\langle F\right\rangle}{M_{eff}}.
\end{equation}
The numerator of the term in the right hand side of the above equation is the deviation of the force from its mean value. We assume that the mirror will fluctuate near the position $x=0$. In this approximation, we can use the static force expression \eqref{force}, i.e. the force when the mirror is staying at the origin, to substitute into the numerator $F-\left\langle F\right\rangle$. In the following we will denote the static force by $F_0$ to avoid confusion with the moving force $F$.

Also note that the factor $\left\langle\dot{q}^2\right\rangle$ in the expression \eqref{fluctuation of force result} for the fluctuation of the static force $F_0$ is logarithmic divergence. And from \eqref{vacuum energy density} we know that the vacuum energy density factor $\left\langle T_{00}\right\rangle$ is $k^2$ divergence, so the fluctuation of $F_0$ is $k^2\ln k$ divergence while the fluctuation of effective mass, which contains the divergent term $\dot{q}^2$, is only $\ln k$ divergence. This implies that the fluctuation of the mirror's position and velocity is mainly determined by the fluctuation of force. So we can further use the mean value of the effective mass $\left\langle M_{eff}\right\rangle$ to substitute $M_{eff}$ to simplify the above equation \eqref{newton eq revised}:
\begin{equation}\label{langevin equation}
\frac{dv}{dt}+\beta v=\frac{F_0}{\left\langle M_{eff}\right\rangle},
\end{equation}
where $v=\frac{dX}{dt}$ is the mirror's velocity, 
\begin{equation}\label{beta}
\beta=\left(\frac{\epsilon^4}{4\pi}\int_0^{+\infty}dk\frac{k^3}{(k^2-\Omega^2)^2+\frac{\epsilon^4}{4}k^2}\right)/\left\langle M_{eff}\right\rangle
\end{equation}
is the damping coefficient,  and we have substituted for $\left\langle F\right\rangle$ by equation \eqref{damping formula}. Both the numerator and denominator of \eqref{beta} contain divergent integrals over $k$. To make precise of the meaning of $\frac{\infty}{\infty}$ type quantities, we first truncate both the integrals by the same high frequency cut-off $k=\Lambda$, and then take $\Lambda$ to infinity. Then the logarithmic divergence of the damping force is magically canceled by the logarithmic divergence of the mirror effective mass, and we get that
\begin{equation}\label{damping coefficient}
\beta=\epsilon^2.
\end{equation}
Equation \eqref{langevin equation} is a Langevin type equation. The solution for the velocity in this equation with initial condition $v(0)=0$ is
\begin{equation}\label{velocity solution}
v(t)=\frac{1}{\left\langle M_{eff}\right\rangle}e^{-\beta t}\int_{0}^{t}{dt'e^{\beta t'}F_0(t')}.
\end{equation}
Then the fluctuation of the velocity is
\begin{equation}\begin{split}
\sigma_{v}(t)=&\left\langle v(t)^2\right\rangle-\left\langle v(t)\right\rangle^2\\
=&\frac{1}{\left\langle M_{eff}\right\rangle^2}e^{-2\beta t}\int_0^t\int_0^tdt_1dt_2e^{\beta(t_1+t_2)}\\
&\quad\quad\quad\quad\quad\quad\cdot Corr(F_0(t_1),F_0(t_2)).
\end{split}\end{equation}
Inserting \eqref{simmaf1f2} into the above expression we get
\begin{equation}\begin{split}
&\sigma_{v}(t)\\
=&\frac{\epsilon^4}{16\pi^2\left\langle M_{eff}\right\rangle^2}\int_{-\infty}^{+\infty}dk|k|\\
\cdot&\int_{-\infty}^{+\infty}dk'\frac{|k'|^3}{(k'^2-\Omega^2)^2+\frac{\epsilon^4}{4}k'^2}\cdot\frac{1}{\beta^2+(|k|+|k'|)^2}\\
\cdot&\left(1-2e^{-\beta t}\cos(|k|+|k'|)t+e^{-2\beta t}\right).
\end{split}\end{equation}

What we are interested is the large time behaviour of the mirror, so now we let $t$ is large enough such that $\beta t\gg 1$, then in such a limit the mirror would be in equilibrium with the quantum scalar field. In this limit, the above expression reduces to
\begin{equation}\label{velocity fluctuation}
\begin{split}
&\sigma_{v}(t)\\
=&\frac{\epsilon^4}{16\pi^2\left\langle M_{eff}\right\rangle^2}\int_{-\infty}^{+\infty}dk|k|\int_{-\infty}^{+\infty}dk'\\
\cdot&\frac{|k'|^3}{(k'^2-\Omega^2)^2+\frac{\epsilon^4}{4}k'^2}\cdot\frac{1}{\beta^2+(|k|+|k'|)^2}.
\end{split}
\end{equation}
The $k'$ integral in \eqref{velocity fluctuation} is convergent and goes as
\begin{equation}
\begin{split}
&\int_{-\infty}^{+\infty}dk'\frac{|k'|^3}{(k'^2-\Omega^2)^2+\frac{\epsilon^4}{4}k'^2}\cdot\frac{1}{\beta^2+(|k|+|k'|)^2}\\
&\quad\quad\quad\sim\frac{2\ln k}{k^2}, \quad\quad\quad as \quad k\to +\infty.
\end{split}
\end{equation}
Therefore the whole integral over $k$, which is divergent, goes as
\begin{equation}
4\int^{\Lambda}dk\frac{\ln k}{k}\sim 2(\ln \Lambda)^2,\quad as \quad \Lambda\to+\infty.
\end{equation}
According to \eqref{effective mass}, we have the pre-factor 
\begin{equation}
\frac{\epsilon^4}{16\pi^2\left\langle M_{eff}\right\rangle^2}\sim 1/(\ln \Lambda)^2,\quad as \quad\Lambda\to+\infty.
\end{equation}
Therefore, after taking the limit $\Lambda\to+\infty$, we obtain
\begin{equation}
\sigma_v=2,
\end{equation}
which means that the standard deviation of the velocity is $\sqrt{2}$ times of light speed! This result is clearly unphysical. It results from our unphysical approximation scheme, namely the small velocity assumption we made in the beginning. In fact, when the velocity of the mirror becomes large, the damping force $\beta v$ would not be linear in velocity and the small velocity approximation is not valid any more. More precisely, from \eqref{jumping force} we know that the damping coefficient is velocity dependent:
\begin{equation}\label{stronger damping force}
\begin{split}
\beta=&\frac{1}{2v}\left((\frac{1+v}{1-v})^{1/2}-(\frac{1-v}{1+v})^{1/2}\right)\epsilon^2\\
=&\left(1+\frac{v^2}{2}+\frac{3v^4}{8}+\frac{5v^6}{16}+...\right)\epsilon^2, 
\end{split}\end{equation}
which reduces to \eqref{damping coefficient} at small velocity approximation.  Therefore, when the mirror's velocity approaches $1$, the damping coefficient would go to infinity to make sure that the mirror's velocity never reach the light speed $1$. If we further fully consider the relativistic effect, the increased mirror's ``relativistic mass" would just make the result even smaller (see the $\gamma^3$ factor in \eqref{final lab frame equation of motion}). Therefore, we can confidently conclude that
\begin{equation}\label{velocity fluctuation bound}
\sigma_v<1,
\end{equation}
which means the mirror's velocity will oscillate wildly due to the fluctuation of quantum field vacuum. However, we will see that this wild oscillation is confined in a small region, that is, the mirror does not diffuse like a Brownian particle.

To prove this, let us calculate the mean squared displacement of the mirror. Strictly speaking, we need to solve the relativistic equation of motion of the mirror. But the relativistic calculation is too messy. Fortunately, we can continue using the non-relativistic Newtonian equation \eqref{langevin equation} to calculate the mean squared displacement. The result is not the true answer but an upper bound of the true answer because when we replace the Newtonian equation \eqref{langevin equation} by relativistic equation of motion \eqref{final lab frame equation of motion}, the mirror would become heavier (due to the $\gamma^3$ factor in \eqref{final lab frame equation of motion}) and the damping force wolud become stronger (see \eqref{stronger damping force}).

Now let us perform the calculation. Integrating \eqref{velocity solution} with time we obtain the solution of the position of the mirror for the initial condition $X(0)=0$ and $v(0)=0$:
\begin{equation}
X(t)=\frac{1}{\left\langle M_{eff}\right\rangle}\int_0^tdt'e^{-\beta t'}\int_0^{t'}dt''e^{\beta t''}F_0(t'').
\end{equation}
Then the mean-squared displacement of the mirror is given by
\begin{equation}\begin{split}
&\sigma_{X}(t)\\
=&\left\langle X(t)^2\right\rangle-\left\langle X(t)\right\rangle^2\\
=&\frac{1}{\left\langle M_{eff}\right\rangle^2}\int_0^tdt_1e^{-\beta t_1}\int_0^{t_1}dt_2e^{\beta t_2}\\
\cdot&\int_0^tdt_3e^{-\beta t_3}\int_0^{t_3}dt_4e^{\beta t_4}Corr(F_0(t_2),F_0(t_4)).
\end{split}\end{equation}
Inserting \eqref{simmaf1f2} into the above expression we get
\begin{equation}\label{mean squared displacement expression}\begin{split}
&\sigma_{X}(t)\\
=&\frac{\epsilon^4}{16\pi^2\left\langle M_{eff}\right\rangle^2}\int_{-\infty}^{+\infty}dk|k|\int_{-\infty}^{+\infty}dk'\\
\cdot&\frac{|k'|^3}{(k'^2-\Omega^2)^2+\frac{\epsilon^4}{4}k'^2}\cdot\frac{1}{\beta^2+(|k|+|k'|)^2}\\
\cdot &\Bigg[\frac{1}{\beta^2}(1-e^{-\beta t})^2+\frac{4\sin^2(\frac{|k|+|k'|}{2}t)}{(|k|+|k'|)^2}\\
&-\frac{1}{\beta}(1-e^{-\beta t})\frac{2\sin(|k|+|k'|)t}{|k|+|k'|}\Bigg].
\end{split}\end{equation}
The double integral of the last two terms over $k$ and $k'$ is convergent, but the effective mass in the denominator is divergent. So the last two terms give no contribution to the mean squared displacement when we take the limit and the above expression reduces to
\begin{equation}\begin{split}
&\sigma_{X}(t)\\
=&\Bigg(\frac{\epsilon^4}{16\pi^2\left\langle M_{eff}\right\rangle^2}\int_{-\infty}^{+\infty}dk|k|\int_{-\infty}^{+\infty}dk'\\
\cdot&\frac{|k'|^3}{(k'^2-\Omega^2)^2+\frac{\epsilon^4}{4}k'^2}\cdot\frac{1}{\beta^2+(|k|+|k'|)^2}\Bigg)\\
&\times\left[\frac{1}{\beta^2}(1-e^{-\beta t})^2\right]
\end{split}\end{equation}

Note that the expression inside the parentheses is just equation \eqref{velocity fluctuation}, the mean squared velocity $\sigma_v$, which is less than $1$ as we concluded in \eqref{velocity fluctuation bound}. Thus, we can further conclude that the mean squared displacement
\begin{equation}\label{mirror mean squared displacement}
\sigma_{X}(t)<\frac{1}{\beta^2}(1-e^{-\beta t})^2,
\end{equation}
or equivalently, the standard deviation of the mirror's position grows with time as
\begin{equation}
\Delta X(t)<\frac{1}{\beta}(1-e^{-\beta t}),
\end{equation}
where $\beta$ is the damping coefficient. 
When $t$ is small, we have
\begin{equation}
\Delta X(t)<t,
\end{equation}
which implies that just after we release the mirror from rest, it starts to diffuse with almost the speed of light! However, as time grows, i.e. when $t\to+\infty$, we always have
\begin{equation}\label{bound}
\Delta X(t)<\frac{1}{\beta},
\end{equation}
which means that the diffusion of the mirror does not continue to increase and its fluctuating motion is confined in the small region $(-\frac{1}{\beta},\frac{1}{\beta})$! The length of this region is inversely proportional  to the damping coefficient, which is physically reasonable because stronger damping would resist the mirror's motion and thus reduce the size of its fluctuating region.

Note also that the damping coefficient $\beta$ is related to the coupling constant $\epsilon$ by Eq.\eqref{damping coefficient} or more precisely by Eq.\eqref{stronger damping force}, which implies that the stronger the coupling, the higher the damping and thus the smaller the range of the fluctuating motion. One might be suspicious of this result since it means that when the coupling $\epsilon$ goes to $0$, the fluctuating range would go to infinity. But if the coupling constant $\epsilon=0$, i.e. there is no interaction between the field and the mirror at all, the mirror should not do any fluctuating motion. It should just sit at the location $x=0$. However, this is not a contradiction but a manifestation of the discontinuity of the expression \eqref{mean squared displacement expression} of the mean squared displacement at $\epsilon=0$. In fact, if $\epsilon=0$, the effective mass $M_{eff}$ reduces to the mirror's finite bare mass $M$ and thus the pre-factor $\frac{\epsilon^4}{16\pi^2\left\langle M_{eff}\right\rangle^2}$ is just $0$, which makes sure that our whole expression \eqref{mean squared displacement expression} is $0$. So our result does satisfy the ``no interaction implies no fluctuation'' requirement.

\section{difference with Brownian motion: the strongly anti-correlation nature of quantum vacuum fluctuations}\label{sec: discussions}
We have concluded in the last section that the fluctuating motion of our mirror would be confined in the small region $(-\frac{1}{\beta},\frac{1}{\beta})$. To better understand the underlying physical mechanism, we would like to compare the fluctuating motion of our mirror with a Brownian particle.

Consider a one dimensional Brownian particle, whose motion is also described by a Langevin type equation:
\begin{equation}\label{Brownian motion langevin equation}
\frac{dv}{dt}+\beta v=\frac{F_B}{m},
\end{equation}
where $m$ is the mass of the Brownian particle, $\beta$ is the damping coefficient and $F_B$ is the stochastic fluctuating force. The only non-trivial difference between the above equation of motion \eqref{Brownian motion langevin equation} for the Brownian particle and the equation of motion \eqref{langevin equation} for our mirror is the different stochastic property of the driven force $F_B$ and $F_0$. 

For the Brownian particle, the force $F_B$ is usually assumed to have a Gaussian probability distribution with correlation function:
\begin{equation}\label{brown correlation}
Corr(F_B(t_1), F_B(t_2))=C\delta(t_1-t_2),
\end{equation}
where $C$ is a constant characterising the strength of the force. The $\delta$-function form of the correlation is an approximation. It means that the force at time $t_1$ is completely uncorrelated with the force at any other time $t_2$. For the motion of a "macroscopic" particle at a much larger time scale compared with the collision time of the molecules, the $\delta$ correlation becomes exact.

However, the force correlation function \eqref{simmaf1f2} for our mirror is quite different. There are two terms in (\ref{simmaf1f2}), each of them is a product of two integrals. The first integral of the first term $\int_{-\infty}^{+\infty}|k|e^{-i|k|(t_1-t_2)}dk$ does not converge under the usual definition of improper integral. However, we can make it converge by analytic continuation, i.e. redefine the integral as
\begin{equation}\begin{split}
f_1(\Delta t)=&\int_{-\infty}^{+\infty}|k|e^{-i|k|(t_1-t_2)}dk\\
=&\lim_{\eta\to 0^+}\int_{-\infty}^{+\infty}|k|e^{-i|k|(t_1-t_2-i\eta)}dk\\
=&-\frac{2}{\Delta t^2},
\end{split}\end{equation}
where $\Delta t=t_1-t_2$. The second integral of the first term
\begin{equation}
f_2(\Delta t)=\int_{-\infty}^{+\infty}\frac{|k'|^3e^{-i|k'|(t_1-t_2)}}{(k'^2-\Omega^2)^2+\frac{\epsilon^4}{4}k'^2}dk'
\end{equation}
conditionally converges to a finite positive value under the usual definition of improper integral. Further, when $\Delta t \to 0$, $f_2(\Delta t)$ logarithmically diverges to $+\infty$. Thus the first term $f_1f_2\to-\infty$ when $\Delta t \to 0$. The second term contains another two integrals, each of them approaches to $0$ when $\Delta t \to 0$. Therefore, due to continuity, the correlation function is always negative for small enough $\Delta t$, and its absolute value can be arbitrarily large, i.e. the force has strong anticorrelation at small time scale. This strong anticorrelation implies that if the force at some time $t_1$ is in positive $x$ direction, after some very short time $\Delta t$, the force would be in the negative $x$ direction. On average, the infinite fluctuations of force at different times are strongly cancelled. This is why although the force fluctuation at any specific instant is infinite, we still obtained the finite fluctuation of the force average in section \ref{sec: force average}. Here it is necessary to point out that Ford and Roman \cite{PhysRevD.72.105010} have also noted and discussed this kind of anti-correlation property of the Minkowski vacuum. Unlike our direct calculations above, they used a sampling function with a characteristic width $a$ to smear out the singularities. The anti-correlations we obtained above agrees with theirs in the limit of $a$ approaches zero. In addition, Parkinson and Ford investigated a related anti-correlation effect in \cite{PhysRevA.84.062102}. 

The fluctuating motion of the Brownian particle and our mirror are different under this two different stochastic fluctuating force. In particular, the mean squared displacement for the Brownian particle grows linearly with time:
\begin{equation}
\sigma_X(t)\sim \frac{C}{\beta^2m^2}t\quad when \quad t\to+\infty,
\end{equation}
which is different from the bounded fluctuating motion of our mirror (see \eqref{bound}). In other words, the Brownian particles would exhibit diffusion while our mirror would be confined in a small region.

\section{discussions and conclusions}\label{sec: conclusions}
We have seen that in our non-gravitational mirror system, the value of quantum vacuum energy does have physical significance in its influence on the fluctuations. It provides an infinite fluctuating force acting on the mirror and gives infinite instantaneous acceleration of the mirror. Astonishingly, this infinity makes sense that, under this fluctuating force, the mirror's fluctuating motion would not diverge but be confined in a small region due to the special properties of vacuum friction and anti-correlation of quantum vacuum fluctuations.

It is clear from the calculations that our mirror does not exhibit Brownian motion and  thus no diffusion happens. Gour and Sriramkumar \cite{Gour:1998my} also studied a mirror interacting with the quantum vacuum using the mirror model \eqref{perfect mirror model} with an artificial high frequency cut-off. However, they concluded that the mirror would experience Brownian motion and thus exhibit diffusion. This conclusion is based on the assumption that ``The stochastic force is completely independent of the position of the Brownian particle" (Page 20 of \cite{Gour:1998my}). This assumption is intrinsically equivalent to the Brownian motion correlation condition \eqref{brown correlation} that we have discussed in the last section. So it is not surprising that this assumption leads to their Brownian motion conclusion. It can be shown by direct calculations that the correlation between the position and the stochastic force is not zero but highly anti-correlated. Following similar procedure we did in this paper, it is not difficult to reproduce the result of a bounded fluctuating motion of the mirror without diffusion.

Jaekel and Reynaud \cite{Jaekel:1992ef} also discussed this issue using an approach based on fluctuation-dissipation theorems. They concluded that a mirror coupled to the Minkowski vacuum would exhibit diffusion which is characterized by a logarithmically increasing behaviour at long times. In addition, Ford etc \cite{PhysRevA.65.062102,Yu:2004tu,Bessa:2008pr}, investigated fluctuating motions of a particle or a mirror in modified quantum vacuums other than the Minkowski vacuum, such as in the presence of boundaries \cite{PhysRevA.65.062102,Yu:2004tu} and in Robertson-Walker Space-Times \cite{Bessa:2008pr}. They also obtained the similar logarithmically increasing quantum diffusion results.

Let us comment on the differences of our results from the works of all of the authors above. The differences mainly come from the fact that (I) we are using different mirror models and thus (II) different methods of handling infinities or singularities. Concretely speaking, the above authors are using the perfectly reflecting mirror model \eqref{perfect mirror model}, which is point-like without any internal structure, by simply imposing a boundary condition. However, a realistic mirror must have some internal structures interacting with the photon field. Our mirror is still point-like but with an internal structure: a internal harmonic oscillator which makes it works like a real mirror. 

This intrinsic difference results in distinct methods of handling infinities. It is well known that treating particles as point-like can result in divergences even in classical field theory, so it is not surprising that they would lead to divergences or singularities. In particular, the authors of \cite{Gour:1998my} and \cite{Jaekel:1992ef} had to treat the infinities by introducing an artificial high frequency cut-off and their results are cut-off dependent; the authors of \cite{PhysRevA.65.062102,Yu:2004tu,Bessa:2008pr} regularized the singularities in the correlation functions by an integration by parts procedure. However, it is not clear what is the correct way to regularize these singular correlation functions to obtain finite results in the point defined limit of ordinary quantum field theory. Unphysical results such as the ``negative" fluctuations were obtained using covariant point separation regularization \cite{Qingdi}. Similar negative mean squared velocity and position fluctuations were also obtained in \cite{PhysRevA.65.062102,Yu:2004tu,Bessa:2008pr} by nonrigorous integration by parts procedure, even though the authors interpreted these results as decreases of uncertainties in position and velocity of quantum particles.

Our mirror model avoids these problems since the infinities disappear naturally even when we take the high frequency cut-off $\Lambda$ to infinity. More precisely, when calculating the fluctuation of the mirror's position, the divergence of the mirror's instantaneous acceleration, which comes from the divergent vacuum energy density, is canceled by vacuum friction and strongly anti-correlated vacuum fluctuations. We are not directly dealing with the actual value of the vacuum energy density, but we find that the infinite value is acceptable in our non-gravitational mirror system in the sense that this infinity only results in finite observable effect. Whether or not the infinite naive expectation value of $T_{00}$ has direct physical effects or could be eliminated by renormalization of the cosmological constant or whether a detailed treatment of the effects of this infinity on the gravitational field could also disappear if one concentrated on observable quantities will be the subject of further work.

\appendix
\section{Derivation of the mirror's equation of motion by directly varying $X(t)$}\label{derive action motion}
We first rewrite the action \eqref{released moving action} as
\begin{equation}
\begin{split}
S&=\frac{1}{2}\iint\left(\left(\frac{\partial\phi}{\partial t}\right)^2-\left(\frac{\partial\phi}{\partial x}\right)^2\right)dt dx\\
&+\int\left(-M+\frac{1}{2}\frac{\dot{q}^2}{1-\dot{X}^2}-\frac{1}{2}\Omega^2q^2\right)\sqrt{1-\dot{X}^2}dt\\
&+\epsilon\int d\left(q\phi\right)-\epsilon\int\dot{q}\phi(t,X(t))dt.
\end{split}
\end{equation}
Varying the above action with respect to the mirror's position $X(t)$ yields
\begin{equation}
\begin{split}
\delta S=&\int\frac{\dot{X}\dot{\delta X}}{\sqrt{1-\dot{X}^2}}\left(M+\frac{1}{2}\left(\frac{dq}{d\tau}\right)^2+\frac{1}{2}\Omega^2q^2\right)dt\\
-&\epsilon\int\dot{q}\phi'(t,X(t))\delta X dt\\
=&-\int\delta X\left[\frac{d}{dt}\left(\gamma M_{eff}\dot{X}\right)+\epsilon\dot{q}\phi'(t,X(t))\right]dt.
\end{split}
\end{equation}
Let $\delta S=0$ we obtain exactly the same equation of motion \eqref{equation of motion in lab frame}.

\section{Derivation of the stress-energy tensor}\label{derive stress-energy tensor}
The stress energy tensor can be determined by the functional derivative of the total action $S$ of the system with respect to the background metric $g_{\mu\nu}$:
\begin{equation}
T^{\mu\nu}=\frac{2}{\sqrt{-g}}\frac{\delta S}{\delta g_{\mu\nu}}.
\end{equation}
To start, let us rewrite the action \eqref{released moving action} in a generic background metric $g_{\mu\nu}$ as follows:
\begin{equation}\begin{split}\label{modified action}
S=&-\frac{1}{2}\iint\sqrt{-g}g^{\mu\nu}\partial_{\mu}\phi\partial_{\nu}\phi dtdx\\
&-M\int\sqrt{-g_{\mu\nu}(t,X(t))dX^{\mu}dX^{\nu}}\\
&+\frac{1}{2}\int\left[(\frac{dq}{d\tau})^2-\Omega^2 q^2\right]d\tau\\
&+\epsilon\int\frac{d\phi}{d\tau}(t(\tau),X(t(\tau)))q(t(\tau))d\tau,
\end{split}
\end{equation}
where $\tau$ is the proper time along the mirror trajectory which is related to the global time coordinate $t$ by
\begin{equation}\label{proper time}
d\tau=\sqrt{-g_{\mu\nu}(t,x)\frac{dX^{\mu}}{dt}\frac{dX^{\nu}}{dt}}dt
\end{equation}
and the last three terms in (\ref{modified action}) are integrated along the mirror trajectory. Here we are using the sign convention $(-,+)$. To obtain the functional derivative, we first change the variable $\tau$ to the global time coordinate $t$ by using (\ref{proper time}) and then transform the first two single integrals in the action (\ref{modified action}) into double integrals, i.e. extend the domain of integration from the line $x=X(t)$ to the whole spacetime, by inserting Dirac delta functions:
\begin{equation}
\begin{split}
&S=-\frac{1}{2}\iint\sqrt{-g}g^{\mu\nu}\partial_{\mu}\phi\partial_{\nu}\phi dtdx\\
-&M\iint\sqrt{-g_{\mu\nu}(t,x)\frac{dX^{\mu}}{dt}\frac{dX^{\nu}}{dt}}\delta(x-X(t))dtdx\\
&+\frac{1}{2}\iint\Bigg[\frac{1}{\sqrt{-g_{\mu\nu}(t,x)\frac{dX^{\mu}}{dt}\frac{dX^{\nu}}{dt}}}(\frac{dq}{dt})^2\\
&-\Omega^2 q^2\sqrt{-g_{\mu\nu}(t,x)\frac{dX^{\mu}}{dt}\frac{dX^{\nu}}{dt}}\Bigg]\\
&\cdot\delta(x-X(t))dtdx+\epsilon\int q(t)d\phi(t,X(t)).
\end{split}
\end{equation}
Varying the above action with respect to $g_{\mu\nu}$ gives
\begin{equation}
\begin{split}\label{action variation}
&\delta S=-\frac{1}{2}\iint\delta(\sqrt{-g})g^{\mu\nu}\partial_{\mu}\phi\partial_{\nu}\phi dtdx\\
&-\frac{1}{2}\iint\sqrt{-g}(\delta g^{\mu\nu})\partial_{\mu}\phi\partial_{\nu}\phi dtdx\\
&-M\iint\frac{(\delta g_{\mu\nu})\frac{dX^{\mu}}{dt}\frac{dX^{\nu}}{dt}}{2\sqrt{-g_{\mu\nu}(t,x)\frac{dX^{\mu}}{dt}\frac{dX^{\nu}}{dt}}}\delta(x-X(t))dtdx\\
&+\frac{1}{2}\iint\left[\frac{1}{-g_{\mu\nu}(t,x)\frac{dX^{\mu}}{dt}\frac{dX^{\nu}}{dt}}(\frac{dq}{dt})^2-\Omega^2 q^2\right]\\
&\cdot\frac{(\delta g_{\mu\nu})\frac{dX^{\mu}}{dt}\frac{dX^{\nu}}{dt}}{2\sqrt{-g_{\mu\nu}(t,x)\frac{dX^{\mu}}{dt}\frac{dX^{\nu}}{dt}}}\delta(x-X(t))dtdx.
\end{split}
\end{equation}
Also, we have
\begin{equation}
\delta(\sqrt{-g})=\frac{1}{2}\sqrt{-g}g^{\mu\nu}\delta g_{\mu\nu}, \delta g^{\mu\nu}=-g^{\mu\lambda}g^{\nu\rho}\delta g_{\lambda\rho}.
\end{equation}
Plugging the above two relations into (\ref{action variation}), we get
\begin{equation}
\begin{split}
&\delta S=\frac{1}{2}\iint\sqrt{-g}\Bigg(\partial^{\mu}\phi\partial^{\nu}\phi-\frac{1}{2}g^{\mu\nu}g^{\lambda\rho}\partial_{\lambda}\phi\partial_{\rho}\phi\\
&+\frac{1}{\sqrt{-g}}\bigg(M+\frac{1}{2}\cdot\frac{1}{-g_{\mu\nu}\frac{dX^{\mu}}{dt}\frac{dX^{\nu}}{dt}}(\frac{dq}{dt})^2+\frac{1}{2}\Omega q^2\bigg)\\
&\quad\quad\cdot\frac{\frac{dX^{\mu}}{dt}\frac{dX^{\nu}}{dt}}{\sqrt{-g_{\mu\nu}\frac{dX^{\mu}}{dt}\frac{dX^{\nu}}{dt}}}\delta(x-X(t)) \Bigg)\delta g_{\mu\nu}dtdx.
\end{split}
\end{equation}
Therefore, the stress energy tensor of the whole system is
\begin{equation}\label{energy momentum tensor}
\begin{split}
&T^{\mu\nu}=\frac{2}{\sqrt{-g}}\frac{\delta S}{\delta g_{\mu\nu}}\\
=&\partial^{\mu}\phi\partial^{\nu}\phi-\frac{1}{2}g^{\mu\nu}g^{\lambda\rho}\partial_{\lambda}\phi\partial_{\rho}\phi\\
+&\frac{1}{\sqrt{-g}}\bigg(M+\frac{1}{2}\cdot\frac{1}{-g_{\mu\nu}\frac{dX^{\mu}}{dt}\frac{dX^{\nu}}{dt}}(\frac{dq}{dt})^2+\frac{1}{2}\Omega q^2\bigg)\\
&\quad\quad\cdot\frac{\frac{dX^{\mu}}{dt}\frac{dX^{\nu}}{dt}}{\sqrt{-g_{\mu\nu}\frac{dX^{\mu}}{dt}\frac{dX^{\nu}}{dt}}}\delta(x-X(t)).
\end{split}
\end{equation}
For the case we are considering, the background metric is flat, i.e. $g_{\mu\nu}=\eta_{\mu\nu}$, then the above expression becomes
\begin{equation}
\begin{split}\label{mirror stress-energy tensor}
T^{\mu\nu}=&\partial^{\mu}\phi\partial^{\nu}\phi-\frac{1}{2}\eta^{\mu\nu}\eta^{\lambda\rho}\partial_{\lambda}\phi\partial_{\rho}\phi\\
+&M_{eff}\frac{\frac{dX^{\mu}}{dt}\frac{dX^{\nu}}{dt}}{\sqrt{1-\dot{X}^2}}\delta(x-X(t)),
\end{split}
\end{equation}
where the effective mass is
\begin{equation}\label{effective mass expression}
M_{eff}=M+\frac{1}{2}\left(\frac{dq}{d\tau}\right)^2+\frac{1}{2}\Omega^2q^2.
\end{equation}

\section*{Acknowledgments}
We thank L.H.Ford for helpful comments on earlier drafts of this paper. We also thank Grigori Volovik for helpful remarks on the estimation of the mirror's effective mass, which triggered us to add a new section to derive the mirror's equation of motion rigorously. W.G.U would like to thank the Canadian Institute for Advanced Research (CIFAR), the Natural Sciences and Engineering Research Council of Canada (NSERC), and the John Templeton Foundation for their support of this research. Qingdi Wang would like to thank UBC for their support of studies during this work through International Partial Tuition Scholarship and Faculty of Science Graduate Award.
\bibliographystyle{unsrt}
\bibliography{vacuum}

\begin{thebibliography}{10}

\bibitem{Martin:2012bt}
Jerome Martin.
\newblock {Everything You Always Wanted To Know About The Cosmological Constant
  Problem (But Were Afraid To Ask)}.
\newblock {\em Comptes Rendus Physique}, 13:566--665, 2012.

\bibitem{scully1997quantum}
M.O. Scully and M.S. Zubairy.
\newblock {\em Quantum Optics}.
\newblock Cambridge University Press, 1997.

\bibitem{Lamb1947}
W.~E. {Lamb} and R.~C. {Retherford}.
\newblock {Fine Structure of the Hydrogen Atom by a Microwave Method}.
\newblock {\em Physical Review}, 72:241--243, August 1947.

\bibitem{Casimir:1948dh}
H.B.G. Casimir.
\newblock {On the Attraction Between Two Perfectly Conducting Plates}.
\newblock {\em Indag.Math.}, 10:261--263, 1948.

\bibitem{Rugh:2000ji}
Svend~Erik Rugh and Henrik Zinkernagel.
\newblock {The Quantum vacuum and the cosmological constant problem}.
\newblock {\em Stud.Hist.Philos.Mod.Phys.}, 2000.

\bibitem{RevModPhys.61.1}
Steven Weinberg.
\newblock The cosmological constant problem.
\newblock {\em Rev. Mod. Phys.}, 61:1--23, Jan 1989.

\bibitem{hobson2006general}
M.P. Hobson, G.~Efstathiou, and A.N. Lasenby.
\newblock {\em General Relativity: An Introduction for Physicists}.
\newblock Cambridge University Press, 2006.

\bibitem{Fulling:1976yv}
S.A. Fulling and P.C.W. Davies.
\newblock {Radiation from a Moving Mirror in Two Dimensional Space-Time:
  Conformal Anomaly}.
\newblock {\em Proc.Roy.Soc.Lond.}, A348:393--414, 1976.

\bibitem{birrell1984quantum}
N.D. Birrell and P.C.W. Davies.
\newblock {\em Quantum Fields in Curved Space}.
\newblock Cambridge Monographs on Mathematical Physics. Cambridge University
  Press, 1984.

\bibitem{Gour:1998my}
Gilad Gour and L.~Sriramkumar.
\newblock {Will small particles exhibit Brownian motion in the quantum vacuum?}
\newblock {\em Found.Phys.}, 29:1917--1949, 1999.

\bibitem{Jaekel:1992ef}
Marc-Thierry Jaekel and Serge Reynaud.
\newblock {Quantum fluctuations of position of a mirror in vacuum}.
\newblock {\em J.Phys.I(France)}, 3:1, 1993.

\bibitem{0305-4470-24-5-014}
G~Barton.
\newblock On the fluctuations of the casimir force.
\newblock {\em Journal of Physics A: Mathematical and General}, 24(5):991,
  1991.

\bibitem{0305-4470-24-23-020}
G~Barton.
\newblock On the fluctuations of the casimir forces. ii. the stress-correlation
  function.
\newblock {\em Journal of Physics A: Mathematical and General}, 24(23):5533,
  1991.

\bibitem{1464-4266-7-3-006}
P~C~W Davies.
\newblock Quantum vacuum friction.
\newblock {\em Journal of Optics B: Quantum and Semiclassical Optics},
  7(3):S40, 2005.

\bibitem{PhysRevD.72.105010}
L.~H. Ford and Thomas~A. Roman.
\newblock Minkowski vacuum stress tensor fluctuations.
\newblock {\em Phys. Rev. D}, 72:105010, Nov 2005.

\bibitem{PhysRevA.84.062102}
Victor Parkinson and L.~H. Ford.
\newblock Model for noncancellation of quantum electric field fluctuations.
\newblock {\em Phys. Rev. A}, 84:062102, Dec 2011.

\bibitem{PhysRevA.65.062102}
Chun-Hsien Wu, Chung-I Kuo, and L.~H. Ford.
\newblock Fluctuations of the retarded van der waals force.
\newblock {\em Phys. Rev. A}, 65:062102, May 2002.

\bibitem{Yu:2004tu}
Hong-wei Yu and L.H. Ford.
\newblock {Vacuum fluctuations and Brownian motion of a charged test particle
  near a reflecting boundary}.
\newblock {\em Phys.Rev.}, D70:065009, 2004.

\bibitem{Bessa:2008pr}
Carlos~H.G. Bessa, Valdir~B. Bezerra, and L.H. Ford.
\newblock {Brownian Motion in Robertson-Walker Space-Times from electromagnetic
  Vacuum Fluctuations}.
\newblock {\em J.Math.Phys.}, 50:062501, 2009.

\bibitem{Qingdi}
Qingdi Wang.
\newblock Black hole fluctuations and negative noise kernel.
\newblock {\em UBC Master Thesis}, 2011.

\end{thebibliography}

\end{document}